\documentclass[11pt]{article} 

\usepackage{amsfonts} 
\usepackage{amssymb} 
\usepackage{amsmath} 
\usepackage{graphicx} 
\usepackage{latexsym}
\usepackage{dcolumn}
\usepackage{subfigure} 
\usepackage{bm}

\newcommand{\opbracket}[3]
        {\langle\hspace{0.5pt}#1\hspace{0.5pt}|\hspace{0.5pt}#2
        \hspace{0.5pt}|\hspace{0.5pt}#3\hspace{0.5pt}\rangle}

\newcommand{\tinymath}[1]{\mbox{\tiny\em #1}}

\newcommand{\be}{\begin{eqnarray}} 
\newcommand{\ee}{\end{eqnarray}} 
\newcommand{\half}{\frac{1}{2}} 
\newcommand{\bra}[1]{ \left \langle #1 \right |} 
\newcommand{\ket}[1]{\left |#1 \right \rangle} 
 
\newcommand{\myvec}[2]{\begin{pmatrix} #1 \\ #2 \end{pmatrix}}
 
\def\k2sum { \sum_{ k \in (-\pi/2,\pi/2) } } 
\def\kpi2sum { \sum_{ k \in (-\pi/2,\pi/2) } } 
\def\kpisigmasum { \sum_{\stackrel { k \in (-\pi,\pi) }{ \sigma = \pm } } } 
\def\kpi2sigmasum { \sum_{\stackrel{k\in(-\pi/2,\pi/2) }{ \sigma = \pm } } }

\def\eqdef{\stackrel{\mbox{\scriptsize def}}{=}}

\begin{document}

\title{\Large\bf One-dimensional quantum walks with absorbing boundaries}

\author{ 
{\sc 
        Eric~Bach\thanks{bach@cs.wisc.edu.}}\\[2mm] 
        Computer Sciences Department, University of Wisconsin\\ 
        1210 W.~Dayton St., Madison, WI~~53706\\[2mm] 
\and 
{\sc 
        Susan Coppersmith\thanks{snc@physics.wisc.edu.},\; 
        Marcel~Paz~Goldschen\thanks{mpgoldschen@students.wisc.edu.},\;} and 
{\sc 
        Robert~Joynt\thanks{rjjoynt@facstaff.wisc.edu.}}\\[2mm] 
        Department of Physics, University of Wisconsin,\\ 
        1150 University Avenue, Madison, WI~~53706\\[2mm] 
\and 
{\sc 
        John~Watrous\thanks{jwatrous@cpsc.ucalgary.ca. }}\\[2mm] 
        Department of Computer Science, University of Calgary,\\ 
        2500 University Drive N.W., Calgary, Alberta, Canada~~T2N 1N4\\[4mm] 
}

\date{\today}

\maketitle 

\thispagestyle{empty}

\begin{abstract} 
In this paper we analyze the behavior of quantum random walks.
In particular we present several new results for the absorption probabilities
in systems with both one and two absorbing walls for the one-dimensional case.
We compute these probabilities both by employing generating functions 
and by use of an eigenfunction approach.
The generating function method is used to determine some simple properties
of the walks we consider, but appears to have limitations.
The eigenfunction approach works by relating the problem of absorption to a
unitary problem that has identical dynamics inside a certain domain,
and can be used to compute several additional interesting properties,
such as the time dependence of absorption.  The eigenfunction
method has the distinct advantage that it can be
extended to arbitrary dimensionality.  We outline the solution
of the absorption probability problem of a (d-1)-dimensional wall
in a d-dimensional space.    
\end{abstract}


\section{Introduction} 
\label{sec:introduction}

Several recent papers have studied the properties of {\em quantum walks}, 
which are quantum computational variants of discrete-time random walks 
\cite{AharonovA+01,AmbainisB+01,Kempe02,KonnoN+02,MackayB+01,Meyer96, 
Meyer01, Meyer02, MooreR01, Watrous01-jcss, YamasakiK+02}. 
The behavior of quantum walks differs from that of ordinary random walks in 
several striking ways, due to the fact that quantum walks exhibit 
interference patterns whereas ordinary random walks do not. 
For instance, mixing times, hitting times, and exit probabilities 
of quantum walks can differ significantly from analogously defined 
random walks \cite{AharonovA+01,AmbainisB+01,Kempe02,MooreR01}.  
One-dimensional quantum walks are also relevant to quantum
chaos~\cite{WojcikD02}.  
(Continuous-time variants of quantum walks have also been proposed and
exhibit significant differences from classical continuous random
walks~\cite{ChildsF+01,FarhiG98}.
However, we will only discuss the discrete-time case in this paper.)

Ordinary random walks have had many applications in computer science, 
particularly as algorithmic tools.
Examples include randomized algorithms for graph connectivity, 2SAT, 
and approximating the permanent
(see, for instance, Lov{\'a}sz~\cite{Lovasz93} for several examples
of algorithmic applications of random walks).
Quantum walks have the potential to offer new tools for the design of quantum 
algorithms, which is one of the primary motivations for studying their 
behavior. 
For example, given that quantum walks on certain simple structures mix faster 
or have faster hitting times than random walks, there is the potential that 
they will admit quantum speed-ups for algorithms based on random walks on
more complicated structures.

In this paper we investigate the behavior of a class of one-dimensional quantum 
walks that are very simple generalizations of the quantum walks introduced 
in Refs.~\cite{Meyer96,AmbainisB+01}. 

One-dimensional quantum walks are not likely to be directly applicable 
to algorithm design.
Although there has been some work on quantum walks on general graphs 
\cite{AharonovA+01}, many questions about quantum walks on general graphs
appear to be quite difficult to answer at the present time.
Thus it is important to extend techniques developed to analyze 
one-dimensional quantum walks to quantum walks on general
graphs.  To this end, we also investigate the behavior of quantum walks
in $d$ dimensions.

Ref.~\cite{AmbainisB+01} demonstrates that one-dimensional quantum walks 
differ qualitatively from classical random walks.
For example, this walk spreads with time as $t$ instead of $\sqrt{t}$. 
Moreover, if the walk is evolved in a system with one absorbing boundary, 
the probability of eventual absorption by the wall is less than unity (in 
contrast to a classical unbiased one-dimensional random walk, where this
probability is unity).
In this paper we generalize and extend the results of
Ref.~\cite{AmbainisB+01} and calculate the absorption probabilities of
one-dimensional quantum walks in systems with one and two absorbing walls. 
We use a combinatorial method as well as an eigenfunction expansion method. 
The combinatorial method is similar to that used in Ref.~\cite{AmbainisB+01}, 
and the eigenfunction method has been used by others to address periodic 
systems~\cite{AmbainisB+01,Meyer01, Meyer02, NayakV00,AharonovA+01,WojcikD02} 
as well as systems with potential steps~\cite{Meyer01, Meyer02}. 
Naturally the two methods agree perfectly in all cases in which results have 
been obtained using both methods.  We extend the eigenfunction method
to general dimensionality.

Our results may be summarized as follows.
First, for one-barrier systems, we obtain exact expressions for the
probability of absorption by the barrier, as a function of the initial
distance to the barrier. 
(Complementing these formulas gives the probability of escape to infinity.) 
These expressions involve integrals of different forms coming from our two
methods of analysis.
Both forms allow asymptotic analysis of the absorption probabilities; in
particular, we compute the limiting probabilities when the initial distance
to the barrier is large. 
We do this both for Hadamard walks and for walks based on more general
unitary transformations.
Next, for the two-barrier Hadamard system, we compute the
long-time limit of the
probability of absorption by each barrier when the walker starts off
very far from one barrier but an arbitrary distance from the other barrier.
Again, the expressions involve integrals whose asymptotic limits
are easily analyzed, so that we can compute the limiting probabilities
when the initial distance to both barriers is large.
We then outline how the eigenfunction method can be used to analyze the
behavior in small systems.
Then we use the eigenfunction method to analyze the time dependence of
the absorption in the limit of long times for walks with both one and
two walls.  We find that the approach to the asymptotic limit is much
slower when there are two walls, where the probability remaining to
be absorbed at time $t$ decays as $1/\sqrt{t}$, than when the system has
one wall, where the probability remaining to be absorbed at time
$t$ is proportional to $1/t^2$.
Finally, we study $d$-dimensional walks, and show that the region over which
most of the probability is distributed by time $t$ has volume proportional to
$t^d$.  We indicate how to solve the problem of the absorption
of a ($d-1$)-dimensional wall without giving explicit results.  

The paper is organized as follows. 
In Section~\ref{sec:definitions} we give definitions for one-dimensional
quantum walks and for the specific processes based on quantum walks that
we consider.
Section~\ref{sec:combinatorial} presents our results based on the combinatorial
approach and Section~\ref{sec:eigenfunction} presents our results based on
the eigenfunction method.
Finally, in Section~\ref{sec:higher_dimensions} we analyze generalizations of
one-dimensional quantum walks to higher dimensions.


\section{Definitions}
\label{sec:definitions}


\subsection{One-dimensional quantum walks}

For any finite or countable set $S$ we may denote by $\mathcal{H}(S)$ 
the Hilbert space of all square-summable functions from $S$ to the 
complex numbers $\mathbb{C}$, along with the usual inner product.
Using the Dirac notation, the space $\mathcal{H}(S)$ has a {\em standard 
basis} $\{\ket{s}\,:\,s\in S\}$, which is orthonormal.
One-dimensional quantum walks are discrete-time quantum processes on 
the space $\mathcal{H}(\mathbb{Z}\times\{L,R\})$.
The standard basis for this space therefore consists of elements of the form 
$\ket{n,d}$, where $n\in\mathbb{Z}$ is the {\em location} and $d\in\{L,R\}$
is the {\em direction} component of such an element.

Given an arbitrary unitary operator $U$ on $\mathcal{H}(\{L,R\})$, define a 
unitary operator $W_{\tinymath{U}}$ acting on
$\mathcal{H}(\mathbb{Z}\times\{L,R\})$ as follows. 
For each standard basis state $\ket{n,d}$ we have 
\[ 
W_{\tinymath{U}}\ket{n,d} = \opbracket{L}{U}{d}\ket{n-1,L} +
\opbracket{R}{U}{d}\ket{n+1,R}, 
\] 
and we extend $W_{\tinymath{U}}$ to all of
$\mathcal{H}(\mathbb{Z}\times\{L,R\})$ by linearity. 
Alternately we may define $W_{\tinymath{U}} = T(I\otimes U)$, where $T$ is
defined by 
\[ 
T\ket{n,L} = \ket{n-1,L},\;\;\;T\ket{n,R} = \ket{n+1,R}, 
\] 
and we identify $\mathcal{H}(\mathbb{Z}\times\{L,R\})$ with the tensor 
product space $\mathcal{H}(\mathbb{Z})\otimes\mathcal{H}(\{L,R\})$ in 
the natural way. 
We use the term {\em one-dimensional quantum walk} to refer generally to any 
process involving the iteration of $W_{\tinymath{U}}$, since such processes are
reminiscent of a particle doing a random walk on a one-dimensional lattice. 

We will also consider quantum walks on higher dimensional lattices
in the final section of this paper---definitions for this type of walk
appear in that section.


\subsection{Absorbing boundaries and exit probabilities}

In this paper we will be interested in the situation in which our system is 
initialized to some state and we alternately apply the operator
$W_{\tinymath{U}}$ and perform some measurement of the system. 
The type of measurements we focus on are as follows.
For each $n\in\mathbb{Z}$, consider the projections $\Pi^n_{\mathrm{yes}}$
and $\Pi^n_{\mathrm{no}}$ defined as 
\[ 
\Pi^n_{\mathrm{yes}} = \ket{n,L}\bra{n,L} + \ket{n,R}\bra{n,R},\;\;\; 
\Pi^n_{\mathrm{no}} = I - \Pi^n_{\mathrm{yes}}. 
\] 
These projections describe a projective measurement that corresponds to 
asking the question ``is the particle at location $n$?''. 
Given a system in state $\ket{\psi}$, the answer is ``yes'' with 
probability $\|\Pi^n_{\mathrm{yes}}\ket{\psi}\|^2$, in which case the state 
of the system becomes
$\Pi^n_{\mathrm{yes}}\ket{\psi}$ (renormalized) and the answer is ``no'' with
probability $\|\Pi^n_{\mathrm{no}}\ket{\psi}\|^2$, in which case the state of
the system becomes $\Pi^n_{\mathrm{no}}\ket{\psi}$ (renormalized).

The first type of process we consider is the one-boundary quantum walk,
which is as follows.
The system being considered is initialized to some state
$\ket{0}(\alpha\ket{L} + \beta\ket{R})$, which corresponds to a particle at
location $0$ and having direction component in state
$(\alpha\ket{L} + \beta\ket{R})$.
Fix an integer $M>0$, which will be the location of our absorbing boundary.
For given $U$, we alternately apply $W_{\tinymath{U}}$ and the measurement
described by $\{\Pi_{yes}^{\tinymath{M}},\Pi_{no}^{\tinymath{M}}\}$, which
gives result ``yes'' if the particle has reached location M and ``no''
otherwise.
The process is repeated until the result ``yes'' is obtained.
The probability that the result ``yes'' is obtained is the
{\em exit probability} for this walk.
For fixed $U$ and given $M$, $\alpha$, and $\beta$, we will denote this exit
probability by $r_{\tinymath{M}}(\alpha,\beta)$.
We also write $p_{\tinymath{M}} = r_{\tinymath{M}}(1,0)$ and
$q_{\tinymath{M}} = r_{\tinymath{M}}(0,1)$ for short, i.e.,
$p_{\tinymath{M}}$ is the exit probability for starting in state $\ket{0,L}$
and $q_{\tinymath{M}}$ is the exit probability for starting in state
$\ket{0,R}$.

We also consider two-boundary quantum walks.
In this case, the particle is initialized in some state
$\ket{0}(\alpha\ket{L}+\beta\ket{R})$ and we alternately apply
$W_{\tinymath{U}}$ and the (commuting) measurements described by 
$\{\Pi_{yes}^{-M_L},\Pi_{no}^{-M_L}\}$ and $\{\Pi_{yes}^{M_R},\Pi_{no}^{M_R}\}$
for $0<M_L,M_R$.
The quantities of interest in this case are the probability of exiting
from the left (i.e., measuring the particle at location $-M_L$)
and the probability of exiting from the right (measuring the particle
at location $M_R$).
Again, in this case the boundaries are absorbing, since the process is
terminated when either measurement gives result ``yes''.

\subsection{Statement of results}

A summary of our results is as follows.

First, we give a complete description of the exit probabilities for
one-boundary quantum walks in one-dimension for arbitrary unitary $U$, any
starting state of the form $\ket{0}(\alpha\ket{L} + \beta\ket{R})$, and any
boundary location $M>0$.
This includes integrals for computing exact exit probabilities, closed
form solutions for the exit probabilities in the limit of large $M$
(one of which proves a recent conjecture of Yamasaki, Kobayashi and Imai
\cite{YamasakiK+02}), and several other results concerning
the behavior of these probabilities.
It is proved that it is enough to consider only real unitary $U$ for the
purposes of analyzing such walks.

Next, for two-boundary Hadamard quantum walks in one dimension,
we present integrals for the exact exit probabilities
when the particle starts out an arbitrary distance from one
wall and an asymptotically large distance from the other.
These integrals are evaluated to yield
closed form solutions for the exit probabilities
when the distance from the first wall is small and when
the particle starts out very far from both walls.
We then calculate the time-dependence of the absorption
probability at long times for
both one- and two-wall walks.

Finally, for $d$-dimensional walks we generalize some of the results
of Ref.~\cite{AmbainisB+01} on one-dimensional walks, including
a derivation of the asymptotic form for the amplitudes associated with the
walk and a demonstration that the amplitude spreads nearly uniformly.
For $d$-dimensional quantum walks with a $(d-1)$-dimensional
barrier an integral for the absorption probability is derived.


\section{Combinatorial analysis} 
\label{sec:combinatorial}


\subsection{Generating functions}

As described in Section~\ref{sec:definitions}, we consider the case where our
system is initialized in some state 
\[
\ket{0}(\alpha\ket{L} + \beta\ket{R})
\]
and we alternately apply the operator $W_{\tinymath{U}}$ and measurement given
by $\{\Pi^{\tinymath{M}}_{\mathrm{yes}},\Pi^{\tinymath{M}}_{\mathrm{no}}\}$
for some $M>0$.
This process is repeated until the measurement gives result ``yes'', 
at which time the process is terminated.

We begin with two special cases:
the first is the case that the starting state is $\ket{0,L}$ and the absorbing
boundary is at $M=1$, and the second is the case that the starting state is
$\ket{0,R}$ and the boundary is at $M=1$.
We will define generating functions for these cases that are used to
determine exit probabilities for all starting states and boundary positions.
For given unitary $U$ define two generating functions $f$ and $g$ as follows: 
\begin{eqnarray*} 
f_{\tinymath{U}}(z) & = & \sum_{t=1}^{\infty} 
\bra{M,R}W_{\tinymath{U}}
(\Pi^{\tinymath{M}}_{\mathrm{no}}W_{\tinymath{U}})^{t-1}\ket{0,L}\,z^t,\\[2mm] 
g_{\tinymath{U}}(z) & = & \sum_{t=1}^{\infty} 
\bra{M,R}W_{\tinymath{U}}(\Pi^{\tinymath{M}}_{\mathrm{no}}
W_{\tinymath{U}})^{t-1}\ket{0,R}\,z^t. 
\end{eqnarray*} 
The coefficient of $z^t$ in $f_{\tinymath{U}}(z)$ is therefore the
(non-normalized) amplitude with which the system is in state $\ket{M,R}$ after
$t$ time steps, assuming the system starts in state $\ket{0,L}$, and
$g_{\tinymath{U}}(z)$ is similar except we start in state $\ket{0,R}$.
We will simply write $f$ and $g$ to denote $f_{\tinymath{U}}$ and
$g_{\tinymath{U}}$ when $U$ is understood.
Thus, for example, the probability that a particle starting in state 
$\ket{0,L}$ is eventually observed at location 0 is 
\[ 
p_1 = \sum_{t\geq 0} \left|[z^t]f(z)\right|^2, 
\] 
where $[z^t]f(z)$ denotes the coefficient of $z^t$ in $f(z)$, and similarly
the probability that a particle starting in state $\ket{0,R}$ is eventually
observed at location 0 is 
\[ 
q_1 = \sum_{t\geq 0} \left|[z^t]g(z)\right|^2. 
\]

The reason that the generating functions $f$ and $g$ are useful for analyzing
exit probabilities for all boundary positions is as follows.
For given $M\geq 2$ consider a generating function defined similarly to
$f$, except for the boundary at location $M$ rather than location 1.
Then this generating function is simply $f(z)(g(z))^{M-1}$, which follows
from the fact that to get from location $0$ to location $M$, the particle
needs to effectively move right $M$ times, and for each move after the
first, the direction component is $R$.
Similarly, the generating function corresponding to starting in state
$\ket{0,R}$ is simply $(g(z))^M$.

Given arbitrary generating functions $u$ and $v$, their {\em Hadamard 
product} is $u\odot v$, defined by 
\[ 
(u\odot v)(z) = \sum_{t\geq 0}([z^t]u(z))([z^t]v(t))\,z^t. 
\] 
Thus, $p_1 = (f\odot\overline{f})(1)$ and $q_1 = (g\odot\overline{g})(1)$.
In general we have 
\begin{equation}
\label{eq:HP-integral}
(u\odot
v)(1)=\frac{1}{2\pi}\int_0^{2\pi}u(e^{i\theta})v(e^{-i\theta})d\theta, 
\end{equation}
provided $\sum_{t\geq 0}([z^t]u(z))([z^t]v(t))$ converges. 
(This follows from results in Section 4.6 of \cite{Titchmarsh39}.)


\subsection{Hadamard Walk}
\label{sec:hadamard_walk}

The most common choice for $U$ in recent papers on one-dimensional quantum
walks has been (or is equivalent to) the following:
\begin{equation}
\label{eq:hadamard}
U \ket{L} = \frac{1}{\sqrt{2}}(\ket{L} + \ket{R}),\hspace*{0.5in} 
U \ket{R} = \frac{1}{\sqrt{2}}(\ket{L} - \ket{R}), 
\end{equation}
i.e., $U$ is the Hadamard transform where we identify $\ket{L} = \ket{0}$ 
and $\ket{R} = \ket{1}$.
The resulting walk has been called the {\em Hadamard walk}.
It turns out that the general behavior of this walk is not specific to the
Hadamard transform, as we will show shortly.
(Nayak and Vishwanath \cite{NayakV00} have also claimed results concerning 
the generality of the Hadamard walk.)
However, it is helpful to first consider the Hadamard transform because
it is simple and we can reduce the behavior of general quantum walks to
the Hadamard walk.

It can be shown that for $U$ as in Eq.~(\ref{eq:hadamard}) we have
\begin{eqnarray*} 
f(z) & = & \frac{1 + z^2 - \sqrt{1 + z^4}}{\sqrt{2}z},\\[2mm] 
g(z) & = & \frac{1 - z^2 - \sqrt{1 + z^4}}{\sqrt{2}z} \:=\:f(z) - \sqrt{2}z. 
\end{eqnarray*} 
We will not argue this here, since later we will derive generating functions 
for arbitrary $U$ that give these generating functions in the case of the
Hadamard transform.

Define $F(\theta) = f(e^{i\theta})$ and $G(\theta) = g(e^{i\theta})$. 
Then by Eq.~(\ref{eq:HP-integral})
\begin{eqnarray*} 
p_1 & = & \frac{1}{2\pi}\int_0^{2\pi}|F(\theta)|^2 d\theta = \frac{2}{\pi}\\ 
q_1 & = & \frac{1}{2\pi}\int_0^{2\pi}|G(\theta)|^2 d\theta = \frac{2}{\pi},
\end{eqnarray*} 
as proved in Ref.~\cite{AmbainisB+01}. 
Thus, in this case a particle starting at location 1 has a 
$1-2/\pi\approx 0.3634$ probability of ``escaping'' the absorbing boundary at 
location 0, which contrasts with the classical unbiased random walk, 
for which the probability of escape is 0.

Suppose now that the boundary is at location $M$ for any $M\geq 1$.
Then from the discussion in the previous subsection we conclude that
\begin{eqnarray} 
p_{\tinymath{M}} & = &
\frac{1}{2\pi}\int_0^{2\pi}|F(\theta)|^2 |G(\theta)|^{2M-2}d\theta 
\label{eq:pnqnint} \\ 
q_{\tinymath{M}} & = &
\frac{1}{2\pi}\int_0^{2\pi}|G(\theta)|^{2M}d\theta. \nonumber 
\end{eqnarray}
These expressions make it very easy to calculate the exit probabilities
in the limit for large $M$.
Since $|G(\theta)| \leq 1$ for all $\theta\in[0,2\pi]$, with strict 
inequality for $\theta\in (\pi/4,3\pi/4) \cup (5\pi/4,7\pi/4)$, we have
\begin{eqnarray*}
\lim_{M\rightarrow\infty} p_{\tinymath{M}} & = &
\lim_{M\rightarrow\infty}
\frac{1}{2\pi}\int_0^{2\pi}|F(\theta)|^2|G(\theta)|^{2M-2}d\theta\\
& = & \frac{1}{2\pi}\int_{-\frac{\pi}{4}}^{\frac{\pi}{4}}
|F(\theta)|^2 d\theta + \frac{1}{2\pi}\int_{\frac{3\pi}{4}}^{\frac{5\pi}{4}} 
|F(\theta)|^2 d\theta\\
& = & \frac{2}{\pi} - \frac{1}{2}\:\approx\:.1366
\end{eqnarray*}
and 
\begin{eqnarray*}
\lim_{M\rightarrow\infty} q_{\tinymath{M}} & = &
\lim_{M\rightarrow\infty} 
\frac{1}{2\pi}\int_0^{2\pi}|G(\theta)|^{2M}d\theta\\
& = &
\frac{1}{2\pi}\int_{-\frac{\pi}{4}}^{\frac{\pi}{4}}d\theta + 
\frac{1}{2\pi}\int_{\frac{3\pi}{4}}^{\frac{5\pi}{4}}d\theta\\
& = & \frac{1}{2}.
\end{eqnarray*}

Now let us calculate the exit probabilities in the limit for large $M$ for
arbitrary directional component in the starting state.
Assume the particle starts in state 
\[
\ket{0}(\alpha\ket{L} + \beta\ket{R}). 
\]
Recall that we denote the exit probability for starting in this state by
$r_{\tinymath{M}}(\alpha,\beta)$. 
The generating function for absorbed paths is now 
\[ 
\alpha f(z)g(z)^{M-1} + \beta g(z)^M = 
(\alpha f(z) + \beta g(z))g(z)^{M-1}. 
\] 
The exit probability is therefore 
\begin{eqnarray*} 
r_{\tinymath{M}}(\alpha,\beta) & = & 
\frac{1}{2\pi}\int_0^{2\pi}|\alpha F(\theta) + \beta G(\theta)|^2 
|G(\theta)|^{2M-2} d\theta\\ 
& \sim & 
\frac{1}{2\pi}\int_{\Gamma} 
|\alpha F(\theta) + \beta G(\theta)|^2 d\theta 
\end{eqnarray*} 
where $\Gamma = (-\frac{\pi}{4},\frac{\pi}{4}) \cup 
(\frac{3\pi}{4},\frac{5\pi}{4})$.
It will be helpful to note that 
\[ 
\frac{1}{2\pi}\int_{\Gamma}F(\theta) G(-\theta) d\theta = 
\frac{1}{2\pi}\int_{\Gamma}|F(\theta)|^2 d\theta - 
\frac{1}{2\pi}\int_{\Gamma}\sqrt{2}e^{-i\theta}F(\theta) d\theta = 
\left(\frac{2}{\pi} - \frac{1}{2}\right) - \frac{1}{\pi} = 
\frac{1}{\pi} - \frac{1}{2}, 
\] 
which follows from the fact that $g(z) = f(z) - \sqrt{2}z$. 
Therefore 
\begin{equation}
\label{eq:rnlimit}
\lim_{M\rightarrow\infty}r_{\tinymath{M}}(\alpha,\beta) 
= 
|\alpha|^2\left(\frac{2}{\pi}-\frac{1}{2}\right) + 
|\beta|^2\left(\frac{1}{2}\right) + 
2\Re(\alpha\overline{\beta})\left(\frac{1}{\pi} - \frac{1}{2}\right). 
\end{equation}
This probability is maximized when $\alpha = \sin(\pi/8)e^{i\phi}$ and 
$\beta = -\cos(\pi/8)e^{i\phi}$, (for arbitrary real $\phi$) giving 
\[ 
\lim_{M\rightarrow\infty}r_{\tinymath{M}}(\alpha,\beta) =
\frac{1}{\sqrt{2}} + \frac{1-\sqrt{2}}{\pi}
\approx 
.5753 
\] 
and is minimized when $\alpha = \sin(3\pi/8)e^{i\phi}$ and 
$\beta = \cos(3\pi/8)e^{i\phi}$, 
giving 
\[
\lim_{M\rightarrow\infty}r_{\tinymath{M}}(\alpha,\beta) =
-\frac{1}{\sqrt{2}} + \frac{1+\sqrt{2}}{\pi} 
\approx 
.0614. 
\]

\subsection{Asymptotics of Exit Probabilities}

It is clear from the integral representation of
$r_{\tinymath{M}}(\alpha,\beta)$ 
and the estimate $|G| \le 1$ that these probabilities decrease with 
$M$, to the limits computed in the last section.
In this subsection we investigate how quickly the limit
$r_\infty = \lim_{\tinymath M \to \infty} r_{\tinymath M}(\alpha,\beta)$
is approached.  We shall prove that
$$
r_{\tinymath M}(\alpha,\beta) = r_\infty + O(M^{-2})
$$
and give an asymptotic series for the remainder.
The principal technique is Watson's lemma \cite{Copson65}, which in this case
reduces to successive integration by parts.

Expanding the integral representation of the last section, we get
\begin{eqnarray*}
r_{\tinymath M}(\alpha,\beta)  
& = &  \frac {|\alpha|^2} {2 \pi} \int_0^{2 \pi} |F|^2 |G|^{2M-2} \; d\theta \\
& + & \frac {|\beta|^2} {2 \pi} \int_0^{2 \pi} |G|^{2M} \; d\theta \\
& + & \frac {2 \Re(\alpha \bar \beta)} {2 \pi} 
\int_0^{2 \pi} \Re \left( F / G \right) |G|^{2M} \; d\theta  .
\end{eqnarray*}
(The last term has no contribution from $\Im(\alpha \bar \beta)$
because of symmetry.)

The coefficient of $|\alpha|^2$ is
$$ 
p_{\tinymath{M}} = \frac 1 {2\pi} \int_0^{2\pi} |F|^2 |G|^{2M-2} \; d\theta 
= \frac 1 {2\pi} \int_0^{2\pi} |F/G|^2 \left( |G|^2 \right)^M \; d\theta. 
$$ 
As a function of $\theta$, the integrand is analytic with the possible
exception of branch points occurring when $\theta = \pm \pi/4, \pm 3\pi/4$. 
Explicitly, when $\pi/4 < \theta < \pi/2$ we have 
$$ 
\left| G \right|^2 
= \frac{( 2\sin \theta - \sqrt{-2 \cos 2\theta})^2} {2} 
$$ 
and 
$$ 
|F/G|^2 = 1 - 2 \cos 2 \theta + 2 \sin \theta \sqrt{-2 \cos 2\theta}, 
$$ 
as can be seen from the formula $|a + b|^2 = |a|^2 + |b|^2 + 2\Re(a\bar b)$.

Making the substitution $e^{-u} = |G(\theta)|^2$ and 
integrating by parts $\nu$ times gives 
$$ 
I_{\tinymath{M}} := \int_{\pi/4}^{\pi/2} |F|^2 |G|^{2M-2} \; d\theta 
= \int_0^{\log(3 + 2\sqrt 2)} \phi(u) e^{- u M} du 
\sim \sum_{k=1}^\nu \frac{\phi^{(k)}(0)}{M^{k+1}}
+ O(1/M^{\nu+2}). 
$$

We now determine the power series for the multiplier $\phi$ 
around $u=0$, as follows. Observe that 
when $\theta = \pi/4 + t$, we have 
$$ 
\frac{( 2\sin \theta - \sqrt{-2 \cos 2\theta})^2}{2} 
= 1 - 2\sqrt 2 s + 4 s^2- 2\sqrt 2 s^3 + \cdots 
$$ 
with $s = t^{1/2}$, whereas 
$$ 
e^{-u} = 1 - u + \frac{u^2}{2} + \cdots \ . 
$$ 
By the implicit function theorem, there is a function $r$, 
analytic around 0, for which $s = r(u)$. 
The coefficients of $r$ can be computed term by term; we have for example 
$$ 
r(u) = \frac{\sqrt 2}{4} u + \frac{\sqrt 2}{96} u^3 + \cdots \ . 
$$ 
Since $r'(u) du = (1/2) t^{-1/2} dt$, we must have 
$$ 
\phi(u) = 2 r(u) r'(u) |F/G|^2_{s = r(u)} . 
$$ 
Coefficients of this may be found using $r$ and the expansion 
of $|F/G|^2$ as a power series in $s$. Explicitly we have 
$$ 
\phi(u) = \frac{1}{4}u + \frac{1}{4}u^2 + \cdots 
$$

The symmetry and periodicity of the integrand imply that 
$$ 
p_{\tinymath{M}} = \left(\frac{2}{\pi} - \frac{1}{2}\right) 
+ 4 \times \frac{1}{2\pi} \int_{\pi/4}^{\pi/2} |F/G|^2 |G|^{2M} d\theta 
$$ 
From this and the above we get the asymptotic series 
\begin{eqnarray}
p_{\tinymath{M}} \sim \left(\frac{2}{\pi} - \frac{1}{2}\right) 
+ \frac{1}{2\pi M^2} 
+ \frac{1}{\pi M^3} 
+ \frac{2}{\pi M^4} 
+ \frac{4}{\pi M^5} 
+ \frac{79}{8 \pi M^6} 
+ \cdots 
\label{eq:pnseries} 
\end{eqnarray} 
Aficionados of the ``law of small numbers'' will appreciate that 
one must develop this series to order 6 to obtain a coefficient 
that does not fit the initial pattern.

Applying similar reasoning, the coefficient of $|\beta|^2$ is
\begin{eqnarray} 
q_{\tinymath{M}} \sim 
\frac{1}{2} 
+ \frac{1}{2\pi M^2} 
+ \frac{1}{2\pi M^4} 
+ \frac{19}{8 \pi M^6} 
+ \cdots \ , 
\label{eq:qnseries} 
\end{eqnarray}
and the coefficient of $2 \Re( \alpha \bar \beta )$ is
\begin{eqnarray}
(pq)_{\tinymath{M}} \sim 
\left(\frac{1}{\pi} - \frac{1}{2}\right)
- \frac 1 {2 \pi M^3}
- \frac 3 {4 \pi M^4}
- \frac 2 {\pi M^5}
- \frac {15} {4 \pi M^6}
+ \cdots \ .
\label{eq:pqnseries} 
\end{eqnarray}


\subsection{Other Transformations}
\label{sec:other}

In this section we argue that the exit probabilities of the Hadamard walk are
not really specific to the Hadamard transform.
The argument can be generalized to other properties of the Hadamard walk.
More generally, we show that it suffices to analyze unitary transformations
$U$ with only real entries.

Suppose instead of using the Hadamard transform we let $U$ be the 
general transformation defined by: 
\begin{equation}
U\ket{L} = a\ket{L} + b\ket{R},\hspace*{0.5in} 
U\ket{R} = c\ket{L} + d\ket{R}. 
\label{eq:generalU}
\end{equation} 
We will consider generating functions $f_{\tinymath{U}}(z)$ and
$g_{\tinymath{U}}(z)$ defined in Section~\ref{sec:hadamard_walk} for this
general transformation $U$.
It is easy to see that these generating functions must satisfy
\begin{eqnarray*} 
f_{\tinymath{U}}(z) & = & b z + a z f_{\tinymath{U}}(z)g_{\tinymath{U}}(z) \\ 
g_{\tinymath{U}}(z) & = & d z + c z f_{\tinymath{U}}(z) g_{\tinymath{U}}(z).
\end{eqnarray*}
Solving these equations for $f_{\tinymath{U}}(z)$ and $g_{\tinymath{U}}(z)$
and taking the solutions that make sense for small powers of $z$ gives 
\begin{eqnarray*} 
f_{\tinymath{U}}(z) & = & 
\frac{1 - (ad - bc)z^2 - 
\sqrt{1-2(ad+bc)z^2+(ad-bc)^2 z^4}} 
{2c z}\\[2mm] 
g_{\tinymath{U}}(z) & = & 
\frac{1+(ad-bc)z^2-\sqrt{1-2(ad+bc)z^2+(ad-bc)^2 z^4}}{2az}. 
\end{eqnarray*} 
The first few terms of these functions are as follows: 
\begin{eqnarray*} 
f_{\tinymath{U}}(z) & = & 
b z + a b d z^3 + 
a b d (a d + b c) z^5 + 
a b d (a^2 d^2 + 3 abcd + b^2c^2) z^7 +\\ 
& & \hspace*{1cm} 
a b d (a^3 d^3 + 6 a b^2 c^2 d + 6 a^2 b c d^2 + b^3 d^3) z^9 + \cdots\\[3mm] 
g_{\tinymath{U}}(z) & = & 
d z + b c d z^3 + 
b c d (a d + b c) z^5 + 
b c d (a^2 d^2 + 3 a b c d + b^2 c^2) z^7 +\\ 
& & \hspace*{1cm} 
b c d (a^3 d^3 + 6 a b^2 c^2 d + 6 a^2 b c d^2 + b^3 d^3) z^9 + \cdots 
\end{eqnarray*} 
Letting $X=ad$ and $Y=bc$, we see that
\begin{eqnarray*} 
f_{\tinymath{U}}(z) & = & 
\frac{1-(X-Y)z^2 - \sqrt{1-2(X+Y)z^2+(X-Y)^2 z^4}}{2 c z}\\[2mm] 
g_{\tinymath{U}}(z) & = & 
\frac{1 + (X - Y)z^2 - \sqrt{1-2(X+Y)z^2+(X-Y)^2 z^4}}{2 a z}. 
\end{eqnarray*}

Next, we can simplify matters by taking into account that $U$ is unitary. 
An arbitrary $2\times 2$ unitary matrix can be written 
\begin{equation}
e^{i\eta} 
\left(
\begin{array}{cc} 
e^{i(\phi + \psi)}\sqrt{\rho}  & e^{i(-\phi + \psi)}\sqrt{1- \rho} \\ 
e^{i(\phi - \psi)}\sqrt{1- \rho} & -e^{i(-\phi - \psi)}\sqrt{\rho}
\end{array} 
\right) 
\label{eq:arbitraryU}
\end{equation}
where $\eta,\phi,\psi$ are real and $0\le\rho\le 1$. 
The global phase of $\eta$ will not affect the behavior of the walk in any 
way, so for simplicity we may set $\eta = 0$ without loss of generality. 
We will write $U_{\rho,\phi,\psi}$ to denote this transformation in order to
stress the dependence on $\rho$, $\phi$, and $\psi$.
This leaves us with 
\begin{eqnarray*} 
X & = & -\rho\\ 
Y & = & 1-\rho. 
\end{eqnarray*} 
Since $X - Y = -1$ and $-(X + Y) = 2\rho-1$, we may write
\begin{eqnarray*} 
f_{U_{\rho,\phi,\psi}}(z) & = & 
\frac{1 + z^2 - \sqrt{1+2(2\rho-1)\,z^2+z^4}}
{2e^{i(-\phi+\psi)}\sqrt{1-\rho} z}
\;\eqdef\; f_{\rho,\phi,\psi}(z)\\[2mm] 
g_{U_{\rho,\phi,\psi}}(z) & = & 
\frac{1 - z^2 - \sqrt{1+2(2\rho-1)\,z^2+z^4}}{2e^{i(\phi+\psi)}\sqrt{\rho} z}
\;\eqdef\; g_{\rho,\phi,\psi}(z).
\end{eqnarray*}
Thus,
\begin{eqnarray*} 
|F_{\rho,\phi,\psi}(\theta)|^2 & = & \frac{1}{4(1-\rho)}\left| 
1 + e^{2i\theta} - \sqrt{1 + 2(2\rho-1) e^{2i\theta} + e^{4i\theta}} 
\right|^2\\[2mm] 
|G_{\rho,\phi,\psi}(\theta)|^2 & = & \frac{1}{4\rho}\left| 
1 - e^{2i\theta} - \sqrt{1 + 2(2\rho-1) e^{2i\theta} + e^{4i\theta}} 
\right|^2,
\end{eqnarray*} 
and as a result we see that, for instance, the quantities 
\begin{eqnarray*} 
p_{\tinymath{M}} & = & 
\frac{1}{2\pi}\int_0^{2\pi}|F_{\rho,\phi,\psi}(\theta)|^2 
|G_{\rho,\phi,\psi}(\theta)|^{2M-2}d\theta\\ 
q_{\tinymath{M}} & = & 
\frac{1}{2\pi}\int_0^{2\pi}|G_{\rho,\phi,\psi}(\theta)|^{2M}d\theta 
\end{eqnarray*} 
depend only on $M$ and $\rho$.
It follows that the exit probabilities for the Hadamard walk would have
been exactly the same had we taken
$U=\frac{1}{\sqrt{2}}\left(\!\!\begin{array}{rr}1&1\\-1&1
\end{array}\!\!\right)$ or $U = \frac{1}{\sqrt{2}}\left(\!\!
\begin{array}{rr}1&i\\i&1\end{array}\!\!\right)$, for instance, rather than
the Hadamard transform.

For arbitrary starting states, the exit probabilities may depend on
$\phi$ and $\psi$ in addition to $\rho$, but this change can be compensated
for by considering slightly different starting states.
Consider the starting state $\ket{0}(\alpha\ket{L}+\beta\ket{R})$ and
transformation $U_{\rho,\phi,\psi}$.
Then the exit probability for this walk is
\[
r_{\tinymath{M}}(\alpha,\beta) = \frac{1}{2\pi}
\int_0^{2\pi}|\alpha F_{\rho,\phi,\psi}(\theta) +
\beta G_{\rho,\phi,\psi}(\theta)|^2 
|G_{\rho,\phi,\psi}(\theta)|^{2M-2}d\theta.
\]
But since $e^{i(-\phi+\psi)}f_{\rho,\phi,\psi} = f_{\rho,0,0}$ and
$e^{i(\phi+\psi)}g_{\rho,\phi,\psi} = g_{\rho,0,0}$,
we see that the exit probability is precisely the same as the exit
probability for the walk given by unitary transformation
$U_{\rho,0,0}$
and starting state
$\alpha e^{i(\phi-\psi)}\ket{0,L} + \beta e^{i(-\phi-\psi)}\ket{0,R}$.

Consequently, it suffices to study the simpler type of transformation 
$U_{\rho,0,0}$, i.e., transformations of the form
\begin{equation}
\label{eq:rho-transformation}
\ket{L} \rightarrow \sqrt{\rho}\ket{L} + \sqrt{1 - \rho}\ket{R}
\;\;\;\mbox{and}\;\;\;
\ket{R} \rightarrow \sqrt{1- \rho}\ket{L} - \sqrt{\rho}\ket{R}
\end{equation}
for $\rho\in[0,1]$ to determine the properties of more general walks.
Such transformations have been considered by Yamasaki, Kobayashi and 
Imai~\cite{YamasakiK+02}. 
In this case we get 
\begin{eqnarray*} 
f_{\tinymath{U}}(z) & = &
\frac{1 + z^2 - \sqrt{1+2(2\rho-1)z^2+ z^4}}{2\sqrt{1-\rho} z}\\[2mm] 
g_{\tinymath{U}}(z) & = & 
\frac{1 - z^2- \sqrt{1+2(2\rho-1)z^2+z^4}}{2\sqrt{\rho} z}. 
\end{eqnarray*} 
Then 
\begin{eqnarray*} 
p_{\tinymath{M}} & = & \frac{1}{2\pi}\int_0^{2\pi}|F_{\tinymath{U}}(\theta)|^2
|G_{\tinymath{U}}(\theta)|^{2M-2}d\theta\\ 
q_{\tinymath{M}} & = & 
\frac{1}{2\pi}\int_0^{2\pi}|G_{\tinymath{U}}(\theta)|^{2M}d\theta. 
\end{eqnarray*}
Letting $\theta_\rho = \frac{\cos^{-1}(1 - 2\rho)} 2$, and 
$\Gamma_\rho = (-\theta_\rho,\theta_\rho) \cup
(\pi - \theta_\rho, \pi + \theta_\rho)$,
we see (after some algebra) that
$|G_{\tinymath{U}}(\theta)| = 1$ for $\theta \in \Gamma_\rho$ and 
$|G_{\tinymath{U}}(\theta)| < 1$ for $\theta \not\in \Gamma_\rho$. 
This gives 
\begin{eqnarray} 
q_{\tinymath{M}} \sim 
\frac 1 {2 \pi} \int_{\Gamma_\rho} d\theta 
= \frac {\cos^{-1}(1 - 2\rho)} \pi 
= \frac {\sin^{-1}(2\rho - 1)} \pi + \frac 1 2. 
\label{eq:qnalimit} 
\end{eqnarray} 
This agrees with Conjecture 1 of \cite{YamasakiK+02} and 
Eq.~(\ref{eq:onewallresult}) below, obtained by the eigenvalue 
method (since $\cos^{-1}(1 - 2\rho) = 2 \sin^{-1} \sqrt \rho$). 
We also have
\begin{eqnarray} 
p_{\tinymath{M}} \sim 
\frac 2 {\pi \sqrt{1/\rho - 1}} 
+ \frac \rho {(1-\rho)\pi} \cos^{-1}(1-2\rho) 
- \frac \rho {(1-\rho)}. 
\label{eq:pnalimit} 
\end{eqnarray} 
This can be computed as $1/(2\pi)\int_{\Gamma_\rho} |F/G|^2 d\theta$,
but we leave this integration job to the interested reader.
This value also follows from Eq.~(\ref{eq:onewallresult}) below.


\section{Eigenfunction method} 
\label{sec:eigenfunction}

In this section we present an eigenfunction method for computing absorption
probabilities of quantum walks. 
We present the method for the quantum random walk introduced
in Ref.~\cite{YamasakiK+02}; the calculation for
the general quantum walk corresponding
to the transformation of Eq.~(\ref{eq:generalU}) is straightforward
but involves slightly more complicated notation.
Letting $L(n,t)$ denote the amplitude of state $\ket{n,L}$ at time $t$
and $R(n,t)$ denote the amplitude of state $\ket{n,R}$ at time $t$, the
dynamical equations are
\be 
\begin{pmatrix} L(n,t) \\ R(n,t) \end{pmatrix} 
=
\begin{pmatrix} \sqrt{\rho} ~L(n+1,t-1) + \sqrt{1-\rho} ~R(n+1,t-1) \nonumber \\
\sqrt{1-\rho} ~L(n-1,t-1) - \sqrt{\rho} ~R(n-1,t-1) \end{pmatrix}~,
\label{eq:qrw} 
\ee 
where we are ignoring boundaries for now.
The Hadamard walk corresponds to the choice
$\rho=1/2$.

Though our main interest here is in studying systems with one or two
absorbing boundaries, it will be very useful to have at our disposal the
eigenfunctions for systems with no boundaries and for periodic systems. 
Therefore, we compute them first.


\subsection{Systems with no boundaries} 

If one looks for solutions of the form 
\be 
\begin{pmatrix} L(n,t) \\ R(n,t) \end{pmatrix} 
= \begin{pmatrix} A_{k} \\ B_{k} \end{pmatrix} e^{i(kn-\omega_{k} t)}~, 
\label{eq:eigenfunctionform}
\ee 
then to satisfy Eq.~(\ref{eq:qrw}) one must have 
\be 
e^{-i\omega_{k}} 
\begin{pmatrix} A_{k} \\ B_{k} \end{pmatrix} 
= {U_{k}} \begin{pmatrix} A_{k} \\ B_{k} \end{pmatrix}~, \nonumber 
\ee 
with 
\be 
{U_{k}} = \begin{pmatrix} 
\sqrt{\rho} ~e^{i{k}} & \sqrt{1-\rho} ~e^{i{k}} \\ 
\sqrt{1-\rho} ~e^{-i{k}} & -\sqrt{\rho} ~e^{-i{k}} 
\end{pmatrix} 
~. 
\nonumber 
\ee 
The characteristic polynomial of $U_{k}$ is 
\be 
\lambda^2 - \sqrt{\rho} ~ \left ( e^{ik}-e^{-ik} \right ) \lambda - 1,
\nonumber 
\ee 
so its eigenvalues are 
\be 
\lambda_{k\pm} =
\sqrt{\rho} \left [i\sin k \pm \sqrt{\cos^2 k + (-1+1/\rho)} \right ]~.
\label{eq:dispersion_relation}
\ee 
Since $U_k$ is unitary we may write $\lambda_{k\pm} = e^{- i \omega_{k\pm}}$
with
\be 
\omega_{k+} & = & -\sin^{-1} \left (\sqrt{\rho} \sin k \right )\\
\omega_{k-} & = & \pi - \omega_{k+}~.
\nonumber 
\ee 
The corresponding eigenfunctions have the form
(\ref{eq:eigenfunctionform}) with
\be 
A_{k\pm} & = & \frac{1}{\sqrt{2N}} 
\sqrt{1\pm\frac{\cos{k}}{\sqrt{ 1/\rho-\sin^2 k }}} 
~, \nonumber \\ 
B_{k\pm} & = & \pm 
\frac{e^{-ik}}{\sqrt{2N}} 
\sqrt{1\mp\frac{\cos{k}}{\sqrt{1/\rho-\sin^2{k}}}}~. 
\label{eq:coefs1} 
\ee 
We have normalized these eigenfunctions so that 
$|A_{k,\sigma}|^2 + |B_{k,\sigma}|^2 = 1/N $, 
which makes the probabilities over any $N$ consecutive lattice 
sites sum to 1. We have also chosen the arbitrary phase to ensure 
that the $A_{k\pm}$ are real and positive.
The group velocity is 
\be 
v_{\pm} \equiv \frac{d\omega_\pm}{dk} 
= \mp\frac{\cos k }{\sqrt{1/\rho-\sin^2 k }} ~; 
\label{eq:group_velocity}
\ee 
$v_-$ is positive for $-\pi/2 < k < \pi/2$, and 
$v_+ = -v_-$ is positive for 
$-\pi < k < -\pi/2$ and for $\pi/2 < k < \pi$.


\subsection{Model with periodic boundary conditions}

We now consider a system whose boundary conditions are periodic with 
period $N$. Any wavefunction of this system can be written as 
a linear superposition of the eigenstates computed above: 
\begin{eqnarray} 
\begin{pmatrix} L(n,t) \\ R(n,t) \end{pmatrix} = 
\kpisigmasum C_{k,\sigma} \left [ 
\begin{pmatrix} 
A_{k,\sigma} \\ B_{k,\sigma} 
\end{pmatrix} 
e^{ikn-i\omega_{k,\sigma}t} 
\right ]~. 
\label{eq:eigenfunction_expansion} 
\end{eqnarray} 
For an initial condition of the form 
\be 
\begin{pmatrix} L(n,0) \\ R(n,0) \end{pmatrix} 
= \delta_{n0} \begin{pmatrix} \alpha \\ \beta \end{pmatrix}~, 
\nonumber 
\ee 
the $C_{k\pm}$ satisfy 
\be 
\begin{pmatrix} A_{k+} & A_{k-} \\ 
B_{k+} & B_{k-} \end{pmatrix} 
\begin{pmatrix} C_{k+} \\ C_{k-} \end{pmatrix} 
= \frac{1}{N} 
\begin{pmatrix} \alpha \\ \beta \end{pmatrix}~. 
\nonumber 
\ee 
Solving for the $C_{k\pm}$ yields 
\be 
C_{k\pm} = A_{k\pm}\alpha + \overline{ B_{k\pm} } \beta~. 
\label{eq:delta_function} 
\ee
In the sequel we will consider periodic systems with $N$ large 
compared to the physical feature of interest (i.e., $M \ll N$), 
with the idea of letting $N \rightarrow \infty$. As an alternative 
to this procedure, one could also average over the continuous variable 
$k$ from the start.


\subsection{Model with one absorbing boundary}

\begin{figure} 
\begin{center}
\includegraphics[height=3cm]{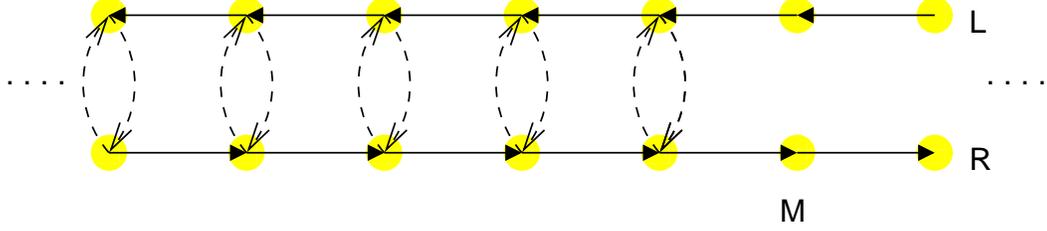} 
\caption{Diagram of system with unitary time evolution 
with dynamics that for all $n<M$ 
are identical to those of a quantum 
walk with an absorbing wall at position $n=M$.} 
\label{fig:onewall} 
\end{center}
\end{figure}

Now consider a system with an absorbing wall 
at location $n=M$, where for definiteness we 
will take $M>0$. 
For our purposes it is useful to think of the 
wall as a boundary 
through which right-movers can be transmitted, 
so that the problem remains 
unitary and probability is conserved. 
We extend the problem so that inside the 
domain the dynamics are identical to the original 
model, and outside the domain 
right-movers move right and left-movers move left, 
as shown in figure~\ref{fig:onewall}. 
Specifically, the model is 
\be 
n<M-1:\qquad \myvec{L(n,t)}{R(n,t)} & = & 
\myvec{\sqrt{\rho} ~L(n+1,t-1)
+      \sqrt{1-\rho} ~R(n+1,t-1)} 
 {\sqrt{1-\rho} ~L(n-1,t-1)
- \sqrt{\rho} ~R(n-1,t-1)} 
\nonumber 
\\ 
n\ge M:\qquad\myvec{L(n,t)}{R(n,t)} & = & 
\myvec{L(n+1,t-1)}{R(n-1,t-1)}~,
\nonumber 
\ee 
together with the boundary condition
\begin{equation}
\myvec{L(M-1,t)}{R(M-1,t)}  =  
\myvec{L(M,t-1)} 
{ \sqrt{1-\rho}~ L(M-2,t-1)-\sqrt{\rho}~ R(M-2,t-1) } ~.
\label{eq:onewallbc} 
\end{equation}
There are no left-movers outside the domain, and 
once the right-movers 
go through the barrier they are no longer converted 
to left-movers, so one must have $L(M,t)=0$ at all 
times $t$. 
The boundary condition Eq.~(\ref{eq:onewallbc}) 
then implies that $L(M-1,t)=0$ also at all times $t$.

\begin{figure}
\begin{center}
\includegraphics[height=6.5cm]{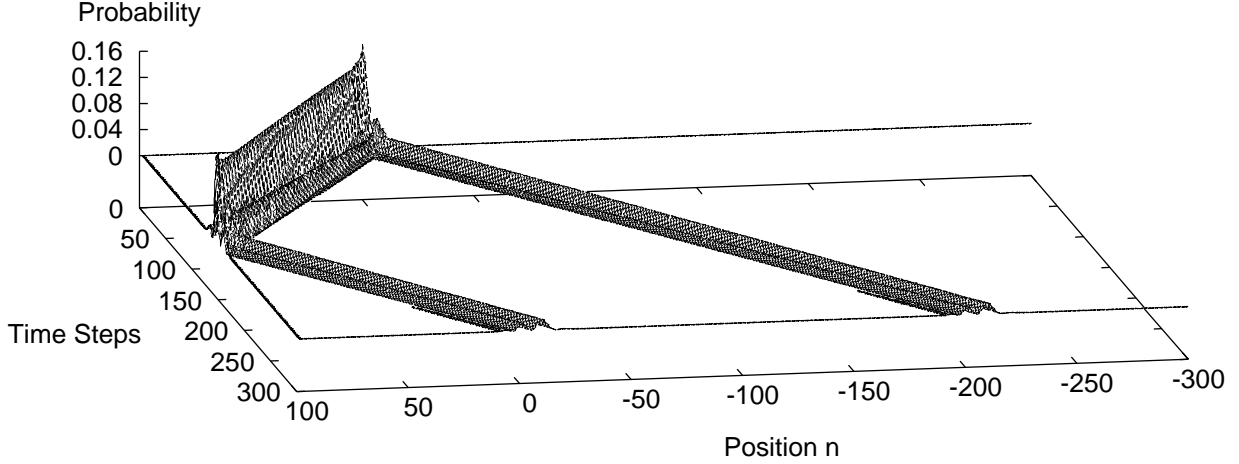} 
\caption{Probability $|L(n,t)|^2+|R(n,t)|^2$
as a function of $n$ and $t$ of a one-dimensional
Hadamard walk with one absorbing wall
started from the initial condition
$L(n,0)=0$ and $R(n,0)\propto \exp(-n^2/100)\cos(\pi x/10)$,
which is overwhelmingly composed of values of $k$
very close to $\pi/10$.
Two wavepackets propagate with group velocities
$\pm v \approx 0.689$
(Eq.~\protect{\ref{eq:group_velocity}};
both $+v$ and $-v$ are seen
because the two bands propagate in opposite directions
for a given $k$);
one reflects from the absorbing boundary at position
$n=100$ with reflection probability $P_r \approx 0.184$
(Eq.~\protect{\ref{eq:reflection_coef}}).
}
\label{figure:reflection}
\end{center}
\end{figure}

Thus, our system 
evolves according to Eq.~(\ref{eq:qrw}) 
and must satisfy 
$L(M-1,t)=0$ at all times $t$. 
The eigenstates of the periodic system discussed 
above clearly do not satisfy this condition. 
The only way that $L(M-1,t)$ can vanish at all times $t$ 
is for contributions from different $k$ that have the 
same value of $\omega$ to interfere destructively 
(otherwise the contributions 
can cancel at some but not all times). 
 From the dispersion relation 
Eq.~(\ref{eq:dispersion_relation}), 
we see that $\omega_{k\pm} = \omega_{(\pi-k)\pm}$, 
and that there are no other degeneracies. 
Thus one is led to look for eigenfunctions of the form 
\be 
\begin{pmatrix} \mathcal{L}_{k\pm}(n,t) \\ 
\mathcal{R}_{k\pm}(n,t) \end{pmatrix} 
= \mathcal{N}_{k\pm}\left [ \myvec{A_{k\pm}}{B_{k\pm}} 
e^{ikn} 
+ \zeta_{k\pm} 
\myvec{A_{(\pi-k)\pm}}{B_{(\pi-k)\pm}} e^{i(\pi-k)n} 
\right ] e^{-i\omega_{k\pm}t}~, 
\label{eq:eigenfunctions1} 
\ee 
where the $\mathcal{N}_{k\pm}$ are normalization constants. 
The coefficients $\zeta_{k\pm}$ are fixed by 
requiring $\mathcal{L}(M-1,t)=0$ for all $t$, or 
\be 
A_{k\pm}e^{ik(M-1)} +\zeta_{k\pm}A_{(\pi-k)\pm}e^{i(\pi-k)(M-1)} = 0~, 
\label{eq:reflection_condition} 
\ee 
yielding (using the $A_{k\pm}$ determined in Eq.~(\ref{eq:coefs1})) 
\be 
\zeta_{k\pm} = -\left (\sqrt{1+\frac{\cos^2 k}{-1+1/\rho} } 
\pm \frac{\cos k}{\sqrt{-1+1/\rho}} \right ) e^{-i(\pi-2k)(M-1)}~. 
\nonumber 
\ee
This situation is analogous to what one finds when one 
considers the scattering of a particle 
by a potential step, as discussed in many elementary quantum 
mechanics texts (see, e.g., \cite{Shankar94}; also see \cite{Meyer02}). 
A rightmoving wave hits the step; the reflected wave 
is leftgoing.
Figure~\ref{figure:reflection}, which shows the time evolution
of a Hadamard walk starting from an initial condition that
is a superposition of a small band of $k$'s,
shows that an absorbing wall indeed acts in this manner.

We know that in the $-$ band the wavefunctions proportional 
to $e^{ikn}$ with wavevectors $k$ in the 
range $(-\pi/2,\pi/2)$ are rightgoing, and we interpret 
with component at wavevector $\pi-k$ 
as the leftgoing piece generated by 
reflection off the boundary. 
The probability that a wave is reflected, 
$P_r(k)$, is just (for $k \in (-\pi/2,\pi/2)$) 
\be 
P_r(k) = |\zeta_{k-}|^2 = \left (\sqrt{1+\frac{\cos^2 k}{-1+1/\rho} } 
- \frac{\cos k}{\sqrt{-1+1/\rho}} \right )^2~. 
\label{eq:reflection_coef1} 
\ee 
Similar calculations for the reflection coefficient of 
rightmoving wavepackets in the other band and also for 
leftmoving waves that are reflected at a boundary 
at which $R(-M+1,t)=0$ at all times $t$ yields that the 
probability of reflection at wavevector $k$ in 
all cases is 
\be 
P_r(k) = |\zeta_{k-}|^2 = \left (\sqrt{1+\frac{\cos^2 k}{-1+1/\rho} } 
- \frac{|\cos k|}{\sqrt{-1+1/\rho}} \right )^2~. 
\label{eq:reflection_coef} 
\ee 
These reflection coefficients agree well with the results 
of our numerical simulations.

We now need to write the initial condition as a superposition 
of eigenfunctions. 
Perhaps the simplest way to do this is to use the 
method of images. 
Thus, we consider a system with no boundary; 
one introduces an image outside the 
domain that is adjusted to enforce the appropriate 
boundary condition, which here is 
$L(M-1,t)=0$ for all $t$. 
We consider initial conditions in which 
the ``physical'' particle is at the origin, 
\be 
\begin{pmatrix} L(n,0)\\R(n,0) \end{pmatrix} 
= \begin{pmatrix} \alpha\\\beta \end{pmatrix} \delta_{n0} ~. 
\nonumber 
\ee 
The form of the unnormalized wavefunction derived above 
suggests that it will be useful to consider together 
the pairs of wavevectors $k, \pi-k$. 
Thus we write 
\be 
\myvec{\alpha}{\beta}\delta_{n,0} = 
\kpi2sigmasum 
\left \{ C_{k,\sigma}\myvec{A_{k,\sigma}}{B_{k,\sigma}}e^{ikn} 
+ C_{\pi-k,\sigma}\myvec{A_{\pi-k,\sigma}}{B_{\pi-k,\sigma}}e^{i(\pi-k)n} 
\right \} ~, 
\nonumber 
\ee 
where the $C_{k,\sigma}$ are given in Eq.~(\ref{eq:delta_function}). 
We now attempt to place an image so that 
the condition $L(M-1,t)=0$ holds at all $t$. 
We guess that the image particle should be 
at $n=2(M-1)$ (again, this is suggested by the form of 
Eq.~(\ref{eq:eigenfunctions1})) and try writing 
\be 
\myvec{L(n,t)}{R(n,t)} = \kpi2sigmasum 
e^{-i\omega_{k,\sigma}t} \left [ 
\left \{ C_{k,\sigma}\myvec{A_{k,\sigma}}{B_{k,\sigma}} 
e^{ikn} 
+ C_{\pi-k,\sigma}\myvec{A_{\pi-k,\sigma}}{B_{\pi-k,\sigma}} 
e^{i(\pi-k)n} 
\right \} \right . \nonumber \\ 
+ \left . \left \{ D_{k,\sigma}\myvec{A_{k,\sigma}}{B_{k,\sigma}} 
e^{ik(n-2(M-1))} 
+ D_{\pi-k,\sigma}\myvec{A_{\pi-k,\sigma}}{B_{\pi-k,\sigma}} 
e^{i(\pi-k)(n-2(M-1))} 
\right \} \right ]~, 
\label{eq:physical_image_wavefunction}
\ee 
where once again we have used the fact that 
$\omega_{k,\sigma} = \omega_{\pi-k,\sigma}$. 
Again, since $L(M-1,t)=0$ for all times, 
we must have for each $k \in (-\pi/2,\pi/2)$: 
\be 
C_{k,\sigma}A_{k,\sigma}e^{ik(M-1)} & + & 
C_{\pi-k,\sigma}A_{\pi-k,\sigma}e^{i(\pi-k)(M-1)} \nonumber \\ 
+~ 
D_{k,\sigma}A_{k,\sigma}e^{-ik(M-1)} & + & 
D_{\pi-k,\sigma}A_{\pi-k,\sigma}e^{-i(\pi-k)(M-1)}=0~. 
\nonumber 
\ee 
It is straightforward to verify that this 
equation is satisfied if we choose 
\be 
D_{k,\sigma} = -e^{i\pi (M-1)} C_{\pi-k,\sigma} 
\frac{A_{\pi-k,\sigma}}{A_{k,\sigma}}~. 
\nonumber 
\ee 
Hence, inside the domain $(n<M)$ 
the time-dependent wavefunction is 
\be 
\myvec{L(n,t)}{R(n,t)} = 
\kpi2sigmasum 
e^{-i\omega_{k,\sigma}t} 
\left \{ 
\myvec{A_{k,\sigma}}{B_{k,\sigma}} e^{ikn} \mathcal{F}_{k,\sigma} 
+ 
\myvec{A_{\pi-k,\sigma}}{B_{\pi-k,\sigma}} e^{i(\pi-k)n} 
\mathcal{G}_{k,\sigma} 
\right \}~, 
\label{eq:wavefunction} 
\ee 
with 
\be 
\mathcal{F}_{k,\sigma} & = & 
\left [ C_{k,\sigma} - e^{i(\pi-2k)(M-1)}C_{\pi-k,\sigma} 
\frac{A_{\pi-k,\sigma}}{A_{k,\sigma}} \right ]\\ 
\mathcal{G}_{k,\sigma} & = & 
\left [ C_{\pi-k,\sigma} - e^{-i(\pi-2k)(M-1)}C_{k,\sigma} 
\frac{A_{k,\sigma}}{A_{\pi-k,\sigma}} \right ]~. 
\label{eq:f_and_g} 
\ee 
As expected, this wavefunction satisfies 
Eq.~(\ref{eq:reflection_condition}).

The wavefunction of Eq.~(\ref{eq:wavefunction}) 
is a superposition of plane waves. 
In the limit of long times, the only components that 
are in the physical domain are the leftgoing waves, 
which for $k\in (-\pi/2,\pi/2)$ are 
($(k,+)$ and $(\pi-k,-)$. 
Therefore, $\Lambda_M$, the probability that the particle escapes 
to $n \rightarrow -\infty$ when the absorbing wall is at $M$, is 
\be 
\Lambda_M = \k2sum |\mathcal{F}_{k,+}(M)|^2 + |\mathcal{G}_{k,-}(M)|^2~. 
\nonumber 
\ee

Using Eqs.~(\ref{eq:f_and_g}),~(\ref{eq:delta_function}), 
and~(\ref{eq:coefs1}), we see that
$\mathcal{G}_{k,-}(M)= \overline{\mathcal{F}_{k,+}(M)}$, so 
\be 
\Lambda_M = 2 \sum_{k \in (-\pi/2, \pi/2)} |\mathcal{F}_{k,+}(M)|^2~, 
\nonumber 
\ee 
with 
\be 
\mathcal{F}_{k,+}(M) = \frac{\alpha}{A_{k+}} \left (A_{k+}^2 
- e^{i(\pi-2k)(M-1)} A_{k-}^2 \right ) 
+ \beta A_{k-}\left ( e^{ik} + e^{i(\pi-2k)(M-1)}e^{-ik} \right )~. 
\nonumber 
\ee 
Converting the sum to an integral, and noting that the integrals 
of terms that are odd in $k$ vanish, we find that 
\be 
\Lambda_M = \frac{1}{\pi} \int_{-\pi/2}^{\pi/2} dk~ 
\mathcal{L}_M(k)~, 
\ee 
with 
\be 
\label{eq:Lofq} 
\mathcal{L}_M(k) & = & \ell^2 \left ( 1 - 
\frac{\cos k }{\sqrt{1/\rho-\sin^2 k}} \right ) 
\left ( \frac{1/\rho + \cos(2k)}{1/\rho-1} 
+ \cos(\pi M)\cos(2k(M-1)) \right ) \nonumber 
\\ 
& + & r^2 \left ( 1-\frac{\cos k }{\sqrt{1/\rho-\sin^2 k }} \right ) 
\left ( 1 - \cos(\pi M) \cos^2 (k M) \right ) \\ 
& + & 
2 \ell r \cos\Phi \left ( \frac{\cos k}{\sqrt{1/\rho-1}} \right ) 
\left (1 - \frac{\cos k}{\sqrt{1/\rho-\sin^2 k}} \right ) 
\left (\cos k - \cos(\pi M)\cos(k(2M-1)) \right )~,\nonumber 
\ee 
where we have written $\alpha = \ell e^{i\Phi_l}$, $\beta = re^{i\Phi_r}$, 
and $\Phi=\phi_l-\phi_r$. 
We obtain 
\be 
\Lambda_M(\rho) & = & \ell^2 \left ( 
-\frac{2}{\pi \sqrt{1/\rho-1}} 
+ \frac{1}{1-\rho} \left ( 1-\frac{2 \rho}{\pi}\sin^{-1}\sqrt{\rho} \right ) 
\nonumber \right .\\ 
& ~ & ~~~~ - \left . \frac{1}{\pi} 
(-1)^M \int_{-\pi/2}^{\pi/2} dk 
\frac{\cos k }{\sqrt{1/\rho-\sin^2 k}} \cos(2k(M-1)) \right ) \nonumber \\ 
& + & 
r^2 \left ( 1-\frac{2}{\pi}\sin^{-1}\sqrt{\rho} 
+ \frac{1}{\pi} (-1)^M \int_{-\pi/2}^{\pi/2} dk 
\frac{\cos k }{\sqrt{1/\rho-\sin^2 k}} \cos(2kM) \right ) 
\label{eq:onewallresult} \\ 
&+& 
\ell r \cos\Phi \left ( 
\frac{1}{\sqrt{1/\rho-1}}-\frac{2}{\pi} 
\left ( 1 + \frac{1-2\rho}{\sqrt{\rho(1-\rho)}}\sin^{-1}\sqrt{\rho} \right ) 
\right . \nonumber \\ 
&~&~~ \left . + 
\frac{2}{\pi} \frac{(-1)^M }{\sqrt{1/\rho-1}}\int_{-\pi/2}^{\pi/2} 
dk \frac{\cos^2 k}{\sqrt{1/\rho-\sin^2 k}} 
\cos(k(2M-1)) 
\right ) 
\nonumber 
~. 
\ee
If instead of Eq.~(\ref{eq:qrw}) we use the more general transformation
of Eq.~(\ref{eq:arbitraryU}),
one finds that the only change in Eq.~(\ref{eq:onewallresult})
is that $\cos\Phi \rightarrow \cos(\Phi+2\phi)$.
Thus, as noted in Sec.~\ref{sec:other},
the absorption probabilities do not depend on $\eta$ or $\psi$, and
the $\phi$ dependence
can be removed by suitable adjustment of the initial condition.

Eq.~(\ref{eq:onewallresult}) only applies when $M>1$, because when 
$M=1$ the 
initial condition is inconsistent with the boundary 
condition unless $\alpha=0$. 
This complication for $M=1$ is easily handled by evolving the walk for 
one time step by hand; for the initial condition 
$\alpha \ket{L,0}+\beta \ket{R,0}$ the probability 
of escape to $-\infty$ when $M=1$ is related to 
$\Lambda_{2L}$, the probability that a walk starting 
in the state $\ket{L,0}$ escapes to $n \rightarrow -\infty$ when the 
wall is at $M=2$, by 
\be 
\Lambda_1 &=& | \alpha\sqrt{\rho} + \beta\sqrt{1-\rho} |^2 \Lambda_{2L}~. 
\nonumber 
\ee 
For the special case of the Hadamard walk ($\rho=1/2$, $\phi=0$),
we find $\Lambda_1 = (1+2\ell r \cos\Phi)(1-2/\pi)$. 
Values for other finite values of $M$ are easily obtained 
by evaluation of the integrals in Eq.~(\ref{eq:onewallresult}); 
a few values for the Hadamard walk
are listed in Table~\ref{tab:probs1}. 
\begin{table} 
\center
\renewcommand{\arraystretch}{1.4}
\begin{tabular}{c|c|c|c} 
M & $\mathcal{C}_\ell$ &$ \mathcal{C}_r$ & $\mathcal{C}_{\ell r}$\\ 
\hline 
1 & $1-\frac{2}{\pi} \approx 0.36338$ 
& $1-\frac{2}{\pi} \approx 0.36338$ 
& $ 2-\frac{4}{\pi} \approx 0.72676$\\
\hline 
2 & $2-\frac{4}{\pi} \approx 0.72676$ 
& $ 3-\frac{8}{\pi} \approx 0.45352$ 
& $ 3-\frac{8}{\pi} \approx 0.45352$\\
\hline 
3 & $4 - \frac{10}{\pi} \approx 0.816901$ 
& $ 13 - \frac{118}{3\pi} \approx 0.479811$ 
& $ 11 - \frac{100}{3\pi} \approx 0.38967$\\
\hline 
4 & 
$ 14 - \frac{124}{3\pi} \approx 0.843191$ 
& $ 65 - \frac{608}{3\pi} \approx 0.489196$ 
& $ 53 - \frac{496}{3\pi} \approx 0.372765$\\
\hline 
5 & 
$ 66-\frac{614}{3\pi} \approx 0.852577$ 
& $ 341-\frac{16046}{15\pi} \approx 0.493304$ 
& $ 277-\frac{13036}{15\pi} \approx 0.367488$\\
\hline
$\infty$ & 
$\frac{3}{2}-\frac{2}{\pi} \approx 0.86338$ 
& $\half$ 
& $1 - \frac{2}{\pi} \approx 0.36338$\\ 
\hline
%
\end{tabular} 
\caption{Coefficients characterizing the probability of 
escape to $-\infty$ of the Hadamard 
walk started in the initial state 
$\alpha \ket{0,L}+\beta \ket{0,R}$ by an absorbing wall at $M$. 
Here, $\alpha = \ell e^{i\phi_l}$, $\beta = r e^{i\phi_r}$, 
and $\Phi=\phi_l-\phi_r$. The quantity $\Lambda_M$, 
the probability of escape to 
$-\infty$ when the wall is located at $M$, is given by 
$\Lambda_M = \ell^2 \mathcal{C}_\ell(M)
+ r^2 \mathcal{C}_r(M) + \ell r \cos\Phi \mathcal{C}_{\ell r}(M)$. 
The probability of absorption by 
the wall is $1-\Lambda_M$. 
} 
\label{tab:probs1} 
\end{table}

In the limit $M \rightarrow \infty$ 
all the $M$-dependent terms vanish 
because of the oscillating integrands. 
The coefficient of the $r^2$ term in the limit $M \rightarrow \infty$ 
has been correctly conjectured by Yamasaki et al.~\cite{YamasakiK+02} 
on the basis of numerical results. 
In the limit $M \rightarrow \infty$, the probability that 
a particle undergoing a Hadamard walk 
escapes to $n \rightarrow -\infty$ is 
\be 
\Lambda_\infty & = & \ell^2 \left ( \frac{3}{2} - \frac{2}{\pi} \right ) 
+ \frac{r^2}{2} + \ell r \cos \Phi \left ( 1-\frac{2}{\pi} \right ) 
~. 
\label{eq:limitescapeprob} 
\ee 
As expected, this agrees with Eq.~(\ref{eq:rnlimit}).
Integration by parts of the integrals in
Eq.~(\ref{eq:onewallresult})
may be used to find
the behavior at large but finite $M$ (c.f.,
Eqs.~(\ref{eq:pnseries}--\ref{eq:pqnseries})).


\subsection{Model with two absorbing boundaries}

Now we discuss the situation in which there are two 
absorbing walls, so the domain of the quantum walk is 
finite. 
We restrict consideration to the Hadamard 
walk ($\rho=1/2$, $\eta=\psi=\phi=0$),
but no essential complications are introduced 
when $\eta$, $\phi$, $\psi$, or $\rho$ take on
different values.
We place the walls at $+M_R$ and $-M_L$ and again 
start the walker out at position $n=0$.

By exploiting 
the results obtained above for the single wall case, it 
is rather simple to compute the absorption probabilities 
when $M_L$ is very large. 
This can be seen by noting that because when the walls are 
very far apart, the evolution can be viewed as a succession 
of reflections off the two walls followed by a single 
transmission. 
 From Eq.~(\ref{eq:reflection_coef}), we know that 
the reflection probability is 
$P_r = (\sqrt{1+\cos^2 k} - |\cos k| )^2$. 
Let us denote by $\mathcal{T}_{LM_R}(k)$ the 
probability of transmission through (or, in other words, 
absorption by) the left boundary of the component at 
wavevector $k$ when the right wall is at $M_R$. 
Because the wavepacket can be 
absorbed at $-M_L$ the first time it hits that boundary, or 
it can be reflected at both the left and right boundaries 
and then absorbed at the left boundary, etc., we have 
\be 
\mathcal{T}_{LM_R}(k) & = & \mathcal{L}_{M_R}(k) \left[ (1-P_r) 
+ P_rP_r(1-P_r) + P_rP_rP_rP_r(1-P_r) + \ldots~ \right] \\ 
& = & \mathcal{L}_{M_R}(k) \frac{1}{1+P_r}~\\ 
& = & \half \mathcal{L}_{M_R}(k) 
\left ( 1+\frac{\cos{k}}{\sqrt{1+\cos^2k}} \right )~, 
\nonumber 
\ee 
where $\mathcal{L}_{M_R}(k)$ is defined in Eq.~(\ref{eq:Lofq}).

When $M_R > 1$, 
the transmission through the left wall when the right wall is 
located at $M_R$, $T_{LM_R}$, is 
thus 
\be 
T_{LM_R} = \frac{1}{\pi} \int_{-\pi/2}^{\pi/2} dk\ \mathcal{T}_{LM_R}(k)~, 
\ee 
with 
\be 
\mathcal{T}_{LM_R}(k) & = & 
\half \ell^2 \frac{1}{1+\cos^2k} 
\left (2+\cos(2k)+\cos(\pi M)\cos(2k(M-1)) \right )\\ 
& ~ &~~~ + \frac{r^2}{2} \frac{1}{1+\cos^2k} (1-\cos(\pi M)\cos(2kM))\\ 
& ~ &~~~ + \ell r\cos\Phi \cos k \frac{1}{1+\cos^2k} 
\left (\cos k - \cos(\pi M)\cos(k(2M-1))\right)~. 
\nonumber 
\ee 
Once again the $M_R=1$ case is done by evolving the system 
by hand for one time step. 
The results for some finite values of $M_R$ are listed in 
Table 2. 
The result for $M_R=1$
was obtained previously in Ref.~\cite{AmbainisB+01}.
\begin{table} 
\center 
\renewcommand{\arraystretch}{1.4}
\begin{tabular}{c|c|c|c} 
M & $D_\ell$ &$ D_r$ & $D_{\ell r}$\\ 
\hline 
1 & $1-\frac{1}{\sqrt{2}} \approx 0.292893$ 
& $1-\frac{1}{\sqrt{2}} \approx 0.292893$ 
& $2-\sqrt{2} \approx 0.585786 $ 
\\
\hline 
2 & $2-\sqrt{2} \approx 0.585786$ 
& $6-4\sqrt{2} \approx 0.343146$ 
& $6-4\sqrt{2} \approx 0.343146$ 
\\
\hline 
3 & $7-\frac{9}{\sqrt{2}} \approx 0.636039$ 
& $35-\frac{49}{\sqrt{2}} \approx 0.351768$ 
& $30-21{\sqrt{2}} \approx 0.301515$ 
\\
\hline 
4 & $36-25\sqrt{2} \approx 0.64461$ 
& $204-144\sqrt{2} \approx 0.353247$ 
& $170-120\sqrt{2} \approx 0.294373$ 
\\
\hline 
5 & $205-\frac{289}{\sqrt{2}} \approx 0.64614$ 
& $ 1189-\frac{1681}{\sqrt{2}} \approx 0.353501$ 
& $ 986-697\sqrt{2} \approx 0.293147$ \\
\hline 
$\infty$ & $1-\frac{1}{2\sqrt{2}} \approx 0.646447$ 
& $\frac{1}{2\sqrt{2}} \approx 0.353553$ 
& $1-\frac{1}{\sqrt{2}} \approx 0.292893$\\ 
\hline
\end{tabular} 
\caption{Coefficients characterizing the probability of 
escape through the left barrier to $-\infty$ of the quantum 
walk started in the initial state 
$\alpha \ket{0,L}+ \beta\ket{0,R}$ with absorbing walls at $M$ 
and at $-M_L$, when $M_L\rightarrow \infty$ 
Here, $\alpha = \ell e^{i\phi_l}$, $\beta = r e^{i\phi_r}$, 
and $\Phi=\phi_l-\phi_r$. The quantity $\mathcal{T}_L(M)$, 
the probability of escape to 
$-\infty$, is given by 
$\mathcal{T}_L(M) = \ell^2 D_\ell(M) + r^2 D_r(M) + \ell r \cos\Phi
D_{lr}(M)$. 
The probability of absorption by 
the right wall at $M$ is $1-\mathcal{T}_L(M)$. 
} 
\label{tab:probs2} 
\end{table}

As $M_R \rightarrow \infty$, the absorption by the 
left wall, $T_{L\infty}$ is 
\be 
T_{L\infty} & = & \ell^2 \left ( 1-\frac{1}{2\sqrt{2}} \right ) 
+ r^2 \left ( \frac{1}{2\sqrt{2}} \right ) 
+ \ell r \cos\Phi \left ( 1-\frac{1}{\sqrt{2}}\right ) 
\nonumber 
\ee 
The minimum value of $T_{L\infty}$ is when $\Phi=\pi$, 
$\ell=\sin( \pi/8)$, 
where $T_{L\infty} =1-1/\sqrt{2}\approx 0.292893$, 
and the maximum value of $T_{L\infty}$ occurs when 
$\Phi=0$, $\ell=\sin(3\pi/8)$, where 
$T_{L\infty} = 1/\sqrt{2} \approx 0.707107$.

When the two walls are close together, the calculations 
are straightforward in principle but tedious in practice. 
The overall strategy of the calculation in this case 
is exactly the same as in all the calculations above---instead of
considering a problem with absorbing barriers, 
one considers a unitary problem whose time evolution is 
identical within the physical domain. 
Here we embed the walk in a large periodic domain;
the model is 
\be 
-M_L+1<n<M_R-1:\qquad \myvec{L(n,t)}{R(n,t)} & = & 
\frac{1}{\sqrt{2}} 
\myvec{L(n+1,t-1) + R(n+1,t-1)}{L(n-1,t-1) - R(n-1,t-1)} 
\nonumber 
\\ 
n\ge M_R~{\rm and}~n\le -M_L:\qquad\myvec{L(n,t)}{R(n,t)} & = & 
\myvec{L(n+1,t-1)}{R(n-1,t-1)} 
\\ 
\myvec{L(M_R-1,t)}{R(M_R-1,t)} & = & 
\myvec{L(M_R,t-1)}{\frac{1}{\sqrt{2}} 
\left (L(M_R-2,t-1)-R(M_R-2,t-1) \right )}~ \nonumber 
\\ 
\myvec{L(-M_L+1,t)}{R(-M_L-1,t)} & = & 
\myvec{\frac{1}{\sqrt{2}}\left (L(-M_L+2,t-1) 
+R(-M_L+2,t-1) \right ) 
} 
{ R(-M_L,t-1) }~. \nonumber 
\label{eq:twowallsbc} 
\ee 
We wish to find the energy eigenstates that satisfy 
\be 
\myvec{L(n,t)}{R(n,t)} = 
\myvec{L_\omega(n)}{R_\omega(n)}e^{-i\omega t}~ 
\nonumber 
\ee 
and are consistent with the boundary conditions 
$L(M_R+N)=L(-M_L)$, $R(M_R+N)=R(-M_L)$.

The eigenvalues and eigenfunctions 
of the model are specified by these conditions 
and can be found similarly to those of systems 
with a finite square well potential \cite{Meyer01}. 
In the ``outer'' region, the eigenfunctions take the 
form 
\be 
\myvec{u_\omega e^{-i\omega n}} 
{v_\omega e^{i\omega n}}e^{-i\omega t}~, 
\nonumber 
\ee 
where the periodic boundary condition is enforced by 
requiring $e^{i\omega( M_L+N)} = e^{-i\omega M_R}$; 
the $u$ and $v$ will be fixed by matching to 
the inner region. 
Since we wish to take $N \rightarrow \infty$, 
there will be a continuum of values of $\omega$. 
Finding the wavefunction for all $M_L$ and $M_R$ is rather 
involved; for every $\omega$ 
one must match the inner and outer wavefunctions, 
keeping in mind that in the inner 
region some of the wavefunctions are of the form 
$e^{\kappa n}$ with $\kappa$ real (this is because 
the quantum walk has a bandgap). 
When the inner region is small, the eigenfunctions 
can be found by brute force matching to the outer 
region. 
For $M_L=1$, $M_R=1$, we have 
$L_\omega(0)=u_\omega$ and $R_\omega(0)=v_\omega$. 
The wavefunction must satisfy the initial condition, 
so $u_\omega=\alpha$ and $v_\omega=\beta$. 
Thus, for this case, as expected, the probability of 
absorption at the right and barriers are $|\beta|^2$ 
and $|\alpha|^2$, respectively. 
When $M_L=1, M_R=2$, 
the coefficients $L_\omega(n)$ in the ``inner'' region satisfy 
\be 
L_\omega(1) & = & u_\omega\nonumber\\ 
R_\omega(1) & = & \frac{1}{\sqrt{2}}e^{i\omega}(L_\omega(0)-R_\omega(0)) 
\nonumber \\ 
L_\omega(0) & = & \frac{1}{\sqrt{2}}e^{i\omega}(L_\omega(1)+R_\omega(2)) 
\nonumber\\ 
R_\omega(0) & = & v_\omega~. 
\ee 
These equations can be solved to yield $L_\omega(0)$ 
and $R_\omega(1)$ in terms of $u_\omega$ 
and $v_\omega$, which 
are in turn fixed by enforcing the initial condition. 
However, the algebra is complicated and will not be given here. 


%
%

\subsection{Arithmetic properties of exit probabilities}

In this subsection we discuss what can be said about the exact
exit probabilities for the Hadamard walk.

Examination of Table 1 above
suggests that the coefficients $\mathcal{C}_\ell(M)$, $\mathcal{C}_r(M)$, 
and $\mathcal{C}_{\ell r} (M)$ are all rational numbers plus
rational multiples of $1/\pi$.  This can be proved as follows.
Taking $\rho = 1/2$ in Eq.~(\ref{eq:onewallresult})
we get for $M>1$
$$
\mathcal{C}_\ell(M)
  = \left(\frac 3 2 - \frac 2 \pi \right)
   - \frac {(-1)^M} \pi 
    \int_{-\pi/2}^{\pi/2}
     \frac {\cos k \ \cos \left( 2(M-1)k  \right)}
     {\sqrt{1 + \cos^2 k}} \; dk,
$$
$$
\mathcal{C}_r(M)
   = \frac 1 2
   + \frac {(-1)^M} \pi 
     \int_{-\pi/2}^{\pi/2}
     \frac {\cos k \  \cos \left( 2Mk \right) }
     {\sqrt{1 + \cos^2 k}} \; dk,
$$
and
$$
\mathcal{C}_{\ell r}(M)
  = 1 - \frac 2 \pi
      + \frac 2 \pi {(-1)^M}
     \int_{-\pi/2}^{\pi/2}
     \frac {\cos^2 k \  \cos \left( (2M-1)k \right) }
     {\sqrt{1 + \cos^2 k}} \; dk.
$$
In the notation of
section 3, these are $1 - p_M$, $1 - q_M$, and  $-2(pq)_M$, 
respectively.

Using the identity $e^{imk} = (\cos k + i \sin k)^m$ and
the binomial theorem, we can express $\cos( mk )$
as a polynomial in $\cos k$.  Furthermore, this polynomial
has only odd powers of $\cos k$ when $m$ is odd, and only even powers
for $m$ even.  Making the substitution $u = \cos k$, we see
(since we started with even integrands)
that each of the integrals is a linear combination, with
rational coefficients, of expressions of the form
$$
    \int_0^1 \frac { u^j } {\sqrt{1 - u^4}} \; du
    =
    \frac {\sqrt \pi} 4  \cdot \frac{\Gamma( (j+1)/4 )} {\Gamma ( (j+3)/4 )}
$$
with $j$ odd.  For such $j$, one of $\{ (j+1)/4, (j+3)/4 \}$ is 
an integer and the other is a half integer.  Thus one of the $\Gamma$
values is an integer and the other is a rational multiple of $\sqrt \pi$,
giving the desired form.

A similar argument can be used to prove that the coefficients
$D_\ell$, $D_r$, and $D_{\ell r}$ of Table 2 are rational
numbers plus rational multiples of $\sqrt 2$.    For this
one needs the integral
$$
\int_0^1 \frac {u^{2j}} {(1 + u^2)\sqrt{1-u^2}} \; du .
$$
If we substitute $u^2 = t$ and use Theorem 2.2.2 of \cite{Andrews99},
we see it equals
$$
\frac {\sqrt \pi} 2
\cdot
\frac {\Gamma (j + 1/2)} {\Gamma(j + 1)} \
{\ _2 F_1 } \left[ {1,\ j + 1/2 \atop j + 1} ; -1 \right],
$$
where ${\ _2 F_1 }$ is the Gaussian hypergeometric function.
It then follows from (15.3.5) and (15.4.11) of \cite{Abramowitz72}
that the coefficients have the desired form.

These observations allow us to generalize a result of \cite{AmbainisB+01}.
Suppose we start both walks at the same distance from the right barrier,
in the state $\alpha = 1$, $\beta = 0$.  Then the escape probability
for the 1-barrier walk is never equal to the limit 
(as the left barrier recedes to $-\infty$) of the corresponding
probability for the 2-barrier walk.  This is so because $\pi$
is not an algebraic number.
The physical interpretation of this result is that no matter how
far away the second wall is, it reflects probability back
to the first wall.

The precise arithmetic properties of these numbers are interesting
to contemplate.  We can show by a combinatorial argument, for example, 
that when $\mathcal{C}_\ell(M) = c_{M,1} + c_{M,2}/\pi$, the denominator of $c_{M,2}$ 
divides the product of the odd numbers in $\{1, \ldots, M-1\}$.  
It also appears that $c_{M,1}$ is always an integer, but we have 
not yet proved this.

%

\subsection{Time dependence of absorption} 
\label{sec:time_dependence}

The eigenfunction method can be used to calculate the 
time evolution of the absorption by the boundaries. 
We consider here the situation when the walker starts
off far from any boundary (M large).  We also specialize
to $\rho=1/2$; once again, generalization to other
transformations is straightforward.

When there are two walls, the approach to the asymptotic 
behavior is quite slow. 
This is because the reflection coefficient tends to unity as 
$k \rightarrow \pi/2$, and the group velocity $v_g$ also 
vanishes as $k \rightarrow \pi/2$. 
Therefore, it takes a very long time for all the 
probability to be absorbed at the walls---the fast-moving 
wavepackets are absorbed quickly, but as the evolution 
proceeds one is left with components that move slower 
and slower and are absorbed less and less. 
The fraction of the probability at wavevector $k$ that 
is absorbed per unit time is 
\[
\frac{\rm [fraction~of~probability~absorbed~ 
per~collision]}{\rm [time~between~collisions]}
\:=\: \frac{1 - \left (\sqrt{1+\cos^2k}-\cos k \right )^2 } 
{2M/v_g}~. 
\]
Thus, we estimate that the probability $\mathcal{P}_k$ 
at wavevector $k$ 
obeys the differential equation 
\be 
\frac{d\mathcal{P}_k}{dt} \sim -\gamma_k \mathcal{P}_k(t)~, 
\ee 
with 
\be 
\gamma_k = \frac{1-(\sqrt{1+\cos^2 k }-\cos k )^2} 
{2M\sqrt{1+\cos^2 k }/\cos k }~,
\ee 
so that $\mathcal{P}_k(t) = \mathcal{P}_k(0)e^{-\gamma_k t}$.
The total probability that 
the particle is inside the region 
at time $t$, $\mathcal{P}(t)$, 
is the integral over wavevectors $k$ of the 
probabilities $\mathcal{P}_k(t)$. 
At long times the probability is dominated by 
$k$'s near $\pm\pi/2$ (since the decay rate tends 
to zero there); expanding for $k$'s near $\pm\pi/2$ 
yields 
\be
\mathcal{P}(t) = \frac{2}{\pi}\int_{0}^{\pi/2}
dk~e^{-(k-\pi/2)^2 t/M}~.
\nonumber
\ee

Now changing variables to $s=(k-\pi/2) \sqrt{t}$ and letting
the limit of integration $\frac{\pi}{2}\sqrt{t}\rightarrow\infty$ gives
\be
\mathcal{P}(t) = \frac{2}{\pi \sqrt{t}}\int_{0}^{\infty}
ds~e^{-s^2 /M}~=\sqrt{\frac{M}{\pi t}}.
\label{eq:t-dep_absorption_2bound}
\ee

\begin{figure}
\begin{center}
\includegraphics[height=7cm]{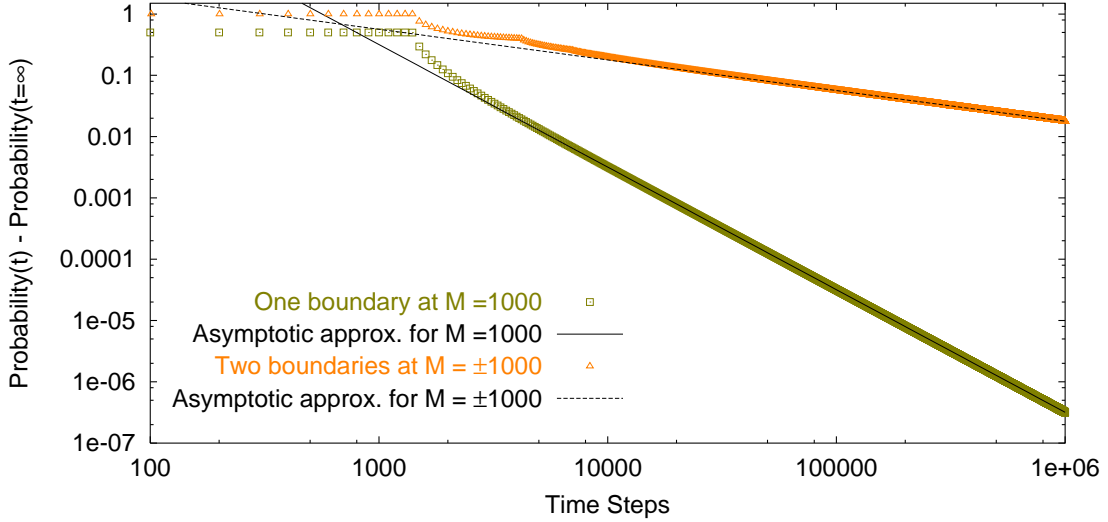}
\caption{The time dependence of $\mathcal{P}(t)$, 
the probability that the particle has not been absorbed, 
is different in systems with one and two absorbing
boundaries. 
With one boundary, 
$\mathcal{P}(t)\propto\mathcal{P}(\infty)+constant/t^2$
(Eq.~(\ref{eq:t-dep_absorption_1bound})), while with two boundaries,
$\mathcal{P}(t)\propto 1/\sqrt{t}$
(Eq.~(\ref{eq:t-dep_absorption_2bound})). 
The faster approach to the asymptotic value for systems with one boundary is
because there are no repeated reflections, while the constant probability
at short times for both systems is because no probability has yet reached the
boundary at position $M$.
Simulations and the analytic forms are all obtained using 
the initial conditions $\ell=0$
and $r=1$. The probability at $t=\infty$ for one boundary is given by 
Eq.~(\ref{eq:onewallresult}), and for two boundaries it is zero.
}
\label{figure:t-dep_absorption}
\end{center}
\end{figure}

When there is one wall that is far from the origin,
the approach of the absorption probability to the
asymptotic value is faster because there is no possibility
of repeated reflections.
The behavior can be characterized 
using Eq.~(\ref{eq:physical_image_wavefunction}) by calculating the
total left-going probability from the physical particle,
\be
\int_{-\pi/2}^{\pi/2}dk~(|C_{k,+}|^2+|C_{\pi-k,-}|^2),
\nonumber
\ee
the fraction of right-going probability from the physical particle
that has yet to reach the wall position $n=M$,
\be
\int_{\cos^{-1}(\frac{M/t}{\sqrt{1-(M/t)^2}})}^{\pi/2}
dk~(|C_{k,+}|^2+|C_{\pi-k,-}|^2),
\nonumber
\ee
and the fraction of left-going probability from the image
particle that has already passed the wall position $n=M$,
\be
\int_{-\cos^{-1}(\frac{M/t}{\sqrt{1-(M/t)^2}})}
^{\cos^{-1}(\frac{M/t}{\sqrt{1-(M/t)^2}})}
dk~(|D_{k,+}|^2+|D_{\pi-k,-}|^2),
\nonumber
\ee
as a function of time. The $C$'s and $D$'s
represent probability amplitudes of plane waves emanating from the
physical and image particle, respectively, as indicated in
Eq.~(\ref{eq:physical_image_wavefunction}), and the limits of integration take into account that
at time $T$, only the $k$'s for which $v_{g} T < M$
have yet to reach the wall. This gives the
asymptotic time dependence for the probability as
\be
\mathcal{P}(t) = \mathcal{P}(\infty) 
+ \frac{M^2}{\pi t^2} + O(\frac{M^3}{t^3}),
\label{eq:t-dep_absorption_1bound}
\ee
where $\mathcal{P}(\infty) = \Lambda_M(\rho=1/2)$ is given 
in Eq.~(\ref{eq:onewallresult}). 
Figure~\ref{figure:t-dep_absorption} demonstrates that the 
asymptotic forms ~(\ref{eq:t-dep_absorption_2bound}) and 
~(\ref{eq:t-dep_absorption_1bound}) provide an excellent 
description of the time dependence for times greater than 
a few times the system size.


\section{Quantum walks in general dimensionality}
\label{sec:higher_dimensions}


\subsection{Introduction}

%
A
quantum walk on a $d$-dimensional lattice
$\vec{n}\in Z^{d}$ may be defined in several ways.
We shall adopt the following.
On each site there are ``internal degrees of freedom'' labeled by an integer
$s\in \lbrack 1,2d]$, where $s$ replaces the $L,R$ notation of the previous
section.
The time-dependent wavefunction is written as $\Psi (\vec{n},s,t)$.
The index $s_{1}$ corresponds to a left move in the first dimension:
$\vec{n}=(n_{1},n_{2}...n_{d})\rightarrow\vec{n}=(n_{1}-1,n_{2}...n_{d})$,
$s_{2}$ corresponds to a right move in the first dimension:
$\vec{n}=(n_{1},n_{2}...n_{d})\rightarrow \vec{n}=(n_{1}+1,n_{2}...n_{d})$,
$s_{3}$ corresponds to a left move in the second dimension:
$\vec{n}=(n_{1},n_{2}...n_{d})\rightarrow \vec{n}=(n_{1},n_{2}-1...n_{d})$,
and so on.
Each time step consists of the ``coin-toss'' step
\[
\Psi (\vec{n},s,t)\rightarrow
\sum_{s^{\prime }=1}^{2d}C_{ss^{\prime }}
\Psi (\vec{n},s^{\prime },t)=\Psi ^{\prime }(\vec{n},s,t),
\]
where $C_{ss^{\prime }}$ is a real $2d\times 2d$ orthogonal matrix, followed
by the walk step
\[
\Psi ^{\prime }(\vec{n},s,t)\rightarrow \sum_{\vec{\delta}}\Psi ^{\prime }
(\vec{n}+\vec{\delta},s_{-\vec{\delta}},t)=\Psi (\vec{n},s,t+1).
\]

Here $\vec{\delta}\in Z^{d}$ is a nearest neighbor separation:
$|\vec{\delta}|=1$.
$\vec{\delta}$ takes on $2d$ possible values.
$s_{\vec{\delta}}$ is the index corresponding to a move in the
$\vec{\delta}$-direction.
This sequence of two successive transformations defines a unitary operator
$U\Psi (\vec{n},s,t)=\Psi (\vec{n},s,t+1)$.

We shall assume that there is a separate coin toss
for each dimension.  This makes $C_{ss^{\prime }}$ into a block-diagonal
matrix consisting of $d$ real $2\times 2$ orthogonal matrices along the diagonal 
and zeroes elsewhere.  A more general definition allows all the elements
of $C_{ss^{\prime }}$ to be nonzero.  Yet a third definition is that adopted by Mackay
{\it et al.} \cite{MackayB+01}.  These authors introduce $d$ qubits at each
site, resulting in a $2^d \times 2^d$ coin-toss matrix.  The second and third 
models can also be
treated by the eigenfunction method, but the results are more cumbersome. 
  
In this section, we give the formal eigenfunction expansion solution for the
asymptotic behavior of our $d$-dimensional quantum walk (first model) and the
survival probability in this model when a $d-1$-dimensional absorbing wall is present.


\subsection{No boundaries}

The problem of finding $\Psi(\vec{n},s,t)=U^{t}\Psi (\vec{n},s,0)$ with
given $\Psi (\vec{n},s,0)$ in the absence of absorbing walls may be solved
by an eigenfunction expansion.
We make the Ansatz 
\[
\Psi (\vec{n},s,t)=\psi _{\vec{q}}(s)e^{i(\vec{q}\cdot \vec{n}-
\omega_{\vec{q}}t)},
\]
where $\psi_{\vec{q}}(s)$ are the eigenfunctions of
$U_{\vec{q}}^{ss^{\prime}}$: 
\begin{equation}
\sum_{s^{\prime }}U_{\vec{q}}^{ss^{\prime }}\psi _{\vec{q}}(s^{\prime})
=e^{-i\omega _{\vec{q}}}\psi _{\vec{q}}(s),  \label{eq:eigenv}
\end{equation}
and the matrix $U_{\vec{q}}^{ss^{\prime }}$ is given by
\[
U_{\vec{q}}^{ss^{\prime }}=\left( 
\begin{array}{ccccc}
C_{11}e^{iq_{1}} & C_{12}e^{iq_{1}} & 0 & 0 & 
... \\ 
C_{21}e^{-iq_{1}} & C_{22}e^{-iq_{1}} & 0 & 0
& ... \\ 
0 & 0 & C_{33}e^{iq_{2}} & C_{34}e^{iq_{2}} & 
... \\ 
0 & 0 & C_{43}e^{-iq_{2}} & C_{44}e^{-iq_{2}}
& ... \\ 
... & ... & ... & ... & ...%
\end{array}%
\right).
\]

The momentum variable $\vec{q}=(q_{1},q_{2},...q_{d})$ lies in the
hypercubic region $\left| q_{i}\right| \leq \pi $, $i=1...d$.
(It is convenient to envision the hypercubic lattice to be extended
periodically in all dimensions, and then taking the period to infinity.)
There are in fact $2d$ solutions to the eigenvalue equations \ref{eq:eigenv}
at a fixed value of $\vec{q}$.
Let us index the eigenvalues $\omega _{\vec{q}}^{\alpha }$ and the
eigenfunctions $\psi _{\vec{q}}^{\alpha }$ by $\alpha =1...2d$.
This band index $\alpha $ replaces the $(+,-)$ labels of the previous
section.

The eigenvalues of the operator $U$ have important degeneracies.

Let us first consider the $2 \times 2$ submatrix $u_1$:
\[
u_1^{ss^{\prime }}(q_1) = \left( 
\begin{array}{cc}
C_{11}e^{iq_{1}} & C_{12}e^{iq_{1}} \\ 
C_{21}e^{-iq_{1}} & C_{22}e^{-iq_{1}} \\ 
\end{array}%
\right).
\]

This is unitary and satisfies
$\det u_1 = \pm1$.
A determinant of $-1$ for a submatrix 
characterizes the Hadamard transformation, but a general
$2 \times 2$ coin-toss operator could also have a positive determinant.
One can show that $u_1$ satisfies the identity
$R u_1(q_1) R^{-1} = det(u_1) u_1(-q_1)$,
where $R$ is the matrix
\[
R = \left( 
\begin{array}{cc}
0 & -1 \\ 
1 & 0 \\ 
\end{array}%
\right).
\]

Thus $u_1(q_1)$ is unitarily equivalent to $u_1(-q_1)$
up to a sign.  If we denote the reflection
$(q_{1},q_{2},...q_{d})\rightarrow (-q_{1},q_{2},...q_{d})$ by
$\vec{q}\rightarrow \vec{q}_{R}$ we have  
$\omega_{\vec{q}}^{\alpha } = \omega_{\vec{q}_{R}}^{\alpha }$ if $\det u_1(\vec{q})=1$
and $\omega_{\vec{q}}^{\alpha }
=\omega _{\pi - q_1, q_2,...q_d)}^{\alpha }$ if $\det u_1(\vec{q}) = - 1$

For a general coin toss matrix $C_{ss^{\prime }}$
no such simple description of the degeneracies exists.
 
A quantum walk that starts at the point $\vec{n}=(0,0,...0)$ with
arbitrary internal state satisfies
\begin{equation}
\Psi (\vec{n},s,t=0)=\delta _{\vec{n},\vec{0}}\Psi _{0}(s)=
\int \sum_{\alpha}a_{\alpha }(\vec{q})\,\psi _{\vec{q}}^{\alpha }(s)\,
e^{i\vec{q}\cdot \vec{n}}Dq,
\label{eq:aalpha}
\end{equation}
so that the coefficients $a_{\alpha }$ are determined by the initial conditions.
We shall employ the shorthand notation
\[
\int f\,Dq\equiv \left( \frac{1}{2\pi }\right) ^{d}\int_{-\pi }^{\pi
}\int_{-\pi }^{\pi }\cdot \cdot \cdot \int_{-\pi }^{\pi
}f\,dq_{1}dq_{2}\cdot \cdot \cdot dq_{d},
\]
and note that
\[
\int e^{i\vec{q}\cdot \vec{n}}Dq=\delta _{\vec{n},\vec{0}}.
\]
Thus, given the initial condition, we find the $a_{\alpha }(\vec{q})$ by
inverting a set of linear equations at each value of $\vec{q}$:
\[
\Psi _{0}(s)=\sum_{\alpha }a_{\alpha }(\vec{q})\,\psi _{\vec{q}}^{\alpha}(s)\,.
\]
The full solution is then
\[
\Psi (\vec{n},s,t)=\sum_{\alpha }\int a_{\alpha }(\vec{q})\,\psi _{\vec{q}
}^{\alpha }(s)\,e^{i(\vec{q}\cdot \vec{n}-\omega _{\vec{q}}^{\alpha }t)}Dq.
\]
To understand the asymptotic properties of this wavefunction at long times,
consider a point that moves at constant velocity $\vec{n}=\vec{c}t$ away
from the origin.
After this substitution, the behavior of the integral as
$t\rightarrow \infty$ may be obtained by the stationary phase method.
The neighborhood(s) of the point or points in $q$-space
$\vec{q}_{i}^{\alpha }(n)$ where
$\nabla _{\vec{q}}\omega _{\vec{q}}^{\alpha }-\vec{c}=0$ dominate
the integral.
The result is
\begin{eqnarray*}
\Psi (\vec{n},s,t) &=&\sum_{\alpha }\int a_{\alpha }(\vec{q})\,
\psi_{\vec{q}}^{\alpha }(s)\,e^{i(\vec{q}\cdot \vec{c}-
\omega_{\vec{q}}^{\alpha })t}\,Dq\\
&\rightarrow& 
\left( \frac{i \pi} {2} \right)^{d/2}
\times \sum_{\alpha }\sum_{i}a_{\alpha }(\vec{q}%
_{i}^{\alpha })\,\psi _{\vec{q}_{i}}^{\alpha }(s)\exp \left[ (i\vec{q}%
_{i}^{\alpha }\cdot \vec{c}-i\omega _{\vec{q}_{i}}^{\alpha })t\right]
\;\left| J_{i}^{\alpha } \right|^{-1/2} t^{-d/2}+{\cal O}(t^{-d/2-1}),
\end{eqnarray*}
where $J_{i}^{\alpha }$ is the Jacobian of the function
$\omega _{\vec{q}}^{\alpha }$ at the point $\vec{q}_{i}^{\alpha }$.
If there are no solutions to the zero-gradient condition, then
the asymptotic behavior of 
$\left| \Psi (\vec{n},s,t)\right|$ is generically determined by
contributions from the boundary of the region of integration
and one finds $\left| \Psi (\vec{n},s,t)\right| = {\cal O} (t^{-d})$.

{}From this expression, we deduce that the probability
$\left| \Psi (\vec{n},s,t)\right| ^{2}$ spreads linearly with time.
The falloff of the wavefunction as $t^{-d/2}$ implies the following physical
picture.
The initial wavepacket is a superposition of waves with various group
velocities
$\vec{v}_{\vec{q}}^{\alpha }=\nabla _{\vec{q}}\omega _{\vec{q}}^{\alpha }$.
Each such component moves according to the ballistic equation
$\vec{n}=\vec{v}_{\vec{q}}^{\alpha }t$.
Because of the limited range of $\vec{q}$, there is a maximum group velocity
in each spatial direction.
This maximum velocity defines the wavefront in that direction.
All components move at constant speed and the overall probability is
normalized: 
\[
\sum_{\vec{n}s}\left| \Psi (\vec{n},s,t)\right| ^{2}\rightarrow t^{-d}\times
\sum_{\left| \vec{n}\right| <t}cst.\sim t^{-d}t^{d}\rightarrow 1.
\]


\subsection{($\mathbf{d-1}$)-dimensional absorbing wall}

We now generalize the absorption problem to a $(d-1)$-dimensional wall located
at $\vec{n}=(M,0,0,...,0)$ with $M>0$.  We shall treat only 
the case where $ \det u_1 = +1$, as the other case has been treated
in detail in one dimension.

In fact, many of the results from Sec.\ref{sec:eigenfunction}
generalize immediately.
We again extend the problem to the full space, stipulating that there is no motion
to the left in the region $n_{1}\geq M$.
Then it is sufficient to solve for the wavefunctions
$\Psi (\vec{n},s,t)=\Psi (n_{1},n_{2},...n_{d},s,t)$ in the region $n_{1}<M$
that satisfy $\Psi (M-1,n_{2},...n_{d},1,t)=0$, which means we have solutions
of the form 
\[
\Psi (\vec{n},s,t)=\sum_{\alpha =1}^{2d}\int_{q_{1}\geq 0}\left[ A_{\alpha }(%
\vec{q},M)\,\psi _{\vec{q}}^{\alpha }(s)\,e^{i\vec{q}\cdot \vec{n}%
}+A_{\alpha }(\vec{q}_R,M)\,\psi _{\vec{q}_{R}}^{\alpha }(s)\,e^{i\vec{q}%
_{R}\cdot \vec{n}}\right] e^{-i\omega _{\vec{q}}^{\alpha }t}Dq.
\]

Enforcing the boundary condition leads to the relation
\[
\sum_{\alpha =1}^{2d}A_{\alpha }(\vec{q}_R,M)\psi _{\vec{q}_{R}}^{\alpha
}(1)=-\sum_{\alpha =1}^{2d}A_{\alpha }(\vec{q},M)\,e^{2iq_{1}(M-1)}\psi _{%
\vec{q}}^{\alpha }(1)
\]
between an expansion coefficient and its reflected counterpart.
As seen already in Sec.\ref{sec:eigenfunction}, this relation is consistent
with the initial conditions, if it is kept in mind that the wavefunction is
only needed in the half-space $n_{1}<M$.
The initial condition is
\[
\Psi (\vec{n},s,t=0)=\delta _{\vec{n},\vec{0}}\Psi _{0}(s)=\sum_{\alpha
=1}^{2d}\int A_{\alpha }(\vec{q},M)\,\psi _{\vec{q}}^{\alpha }(s)\,e^{i\vec{q%
}\cdot \vec{n}}Dq.
\]
Using the method of images or otherwise, we invert this expression to
determine the $A_{\alpha }(\vec{q},M)$ and the final solution for $n_{1}<M$
is
\[
\Psi (\vec{n},s,t)=\sum_{\alpha =1}^{2d}\int A_{\alpha }(\vec{q},M)\psi _{%
\vec{q}}^{\alpha }(s)e^{i\vec{q}\cdot \vec{n}}e^{-i\omega _{q}^{\alpha }t}Dq.
\]
To obtain the survival probability $\Lambda _{M}$, we note that at long
times only leftmoving waves $(v_{1}^{\alpha }(\vec{q})=\partial 
\omega _{\vec{q}}^{\alpha }/\partial q_{1}<0)$ will be present in the
physical domain:
\[
\Lambda _{M}=\sum_{\alpha =1}^{2d}\sum_{s=1}^{2d}\int_{v_{1}^{\alpha }(\vec{q%
})<0} \left| A_{\alpha }(\vec{q},M)\psi _{\vec{q}}^{\alpha
}(s)\right| ^{2}Dq.
\]

This expression manifestly satisfies $0 < \Lambda_M <1$, except for special
choices of the initial condition.  This is in sharp distinction to the classical
random walk, for which the survival probability vanishes for all $d$ and $M$. 



\subsection*{Acknowledgments} 

E.B., S.N.C., M.G., and R.J.\ gratefully acknowledge support from the 
U.S.~NSF QuBIC program, award number 013040. 
J.W.'s research was supported by Canada's NSERC.


\bibliographystyle{plain} 
\bibliography{walks}


\end{document}